\documentclass[conference]{IEEEtran}

\usepackage{float}
\usepackage{xcolor}
\usepackage{mathtools}
\usepackage{enumerate}
\usepackage{multirow}
\usepackage{amssymb}
\usepackage{amsmath}
\usepackage{eqnarray}
\usepackage{booktabs}
\usepackage{array}
\usepackage[]{algorithm}
\usepackage{clrscode3e}
\usepackage[pdftex]{graphicx}

\definecolor{zblack}{HTML}{000000}
\definecolor{zblue}{HTML}{0072BD}
\definecolor{zred}{HTML}{D85319}
\definecolor{zyellow}{HTML}{BC9020}
\definecolor{zpurple}{HTML}{7E2F8E}
\definecolor{zgreen}{HTML}{77AC30}

\usepackage[font={small}]{caption,subfig}  
\DeclarePairedDelimiter{\floor}{\lfloor}{\rfloor}
\DeclarePairedDelimiter{\angled}{\langle}{\rangle}

\begin{document}



\title{ZNN\emph{i} -- Maximizing the Inference Throughput of 3D
  Convolutional Networks on Multi-Core CPUs and GPUs}

\author{\IEEEauthorblockN{Aleksandar Zlateski\IEEEauthorrefmark{1},
   Kisuk Lee\IEEEauthorrefmark{2}}
 \IEEEauthorblockA{\IEEEauthorrefmark{1}Electrical Engineering and
   Computer Science Dept.\\ \IEEEauthorrefmark{2}Brain and
   Cognitive Sciences Dept.\\ Massachusetts Institute of
   Technology\\ Cambridge, MA 02139 USA\\ \IEEEauthorrefmark{1}{\tt
     zlateski@mit.edu}, \IEEEauthorrefmark{2}{\tt kisuklee@mit.edu}}
 \and \IEEEauthorblockN{H. Sebastian Seung}
 \IEEEauthorblockA{Neuroscience Institute\\ and Computer
   Science Dept.\\ Princeton University\\ Princeton, NJ 08540
   USA\\ {\tt sseung@princeton.edu} }}

\maketitle


\begin{abstract}

  Sliding window convolutional networks (ConvNets) have become a
  popular approach to computer vision problems such as image
  segmentation, and object detection and localization. Here we
  consider the problem of inference, the application of a previously
  trained ConvNet, with emphasis on 3D images.  Our goal is to
  maximize throughput, defined as average number of output voxels
  computed per unit time.  Other things being equal, processing a
  larger image tends to increase throughput, because fractionally less
  computation is wasted on the borders of the image.  It follows that
  an apparently slower algorithm may end up having higher throughput
  if it can process a larger image within the constraint of the
  available RAM.  We introduce novel CPU and GPU primitives for
  convolutional and pooling layers, which are designed to minimize
  memory overhead. The primitives include convolution based on highly
  efficient pruned FFTs. Our theoretical analyses and empirical tests
  reveal a number of interesting findings.  For some ConvNet
  architectures, cuDNN is outperformed by our FFT-based GPU
  primitives, and these in turn can be outperformed by our CPU
  primitives. The CPU manages to achieve higher throughput because of
  its fast access to more RAM.  A novel primitive in which the GPU
  accesses host RAM can significantly increase GPU throughput.
  Finally, a CPU-GPU algorithm achieves the greatest throughput of
  all, $10\times$ or more than other publicly available
  implementations of sliding window 3D ConvNets.  All of our code has
  been made available as open source project.

\end{abstract}


\section{Introduction}

  Waving the banner of ``deep learning,'' researchers have revived the
  use of convolutional networks (ConvNets) for computer vision.  The
  revival has been driven by increases in the speed of ConvNet
  training made possible by GPU
  implementations~\cite{chellapilla2006high, scherer2010accelerating,
  strigl2010performance, ciresan2011flexible}. Here we are concerned
  with the problem of inference, applying a previously trained ConvNet
  to a new image.  Fast inference is critical for big data
  applications involving large numbers of images and/or very large
  images. Billions of photos and millions of videos are shared online
  every day~\cite{MeekerReport14, ReelSEO}.  Scientists are also
  generating large amounts of image data.  For example, high-speed
  electron microscopy can generate a petascale 3D image from a cubic
  millimeter of brain in a few weeks~\cite{Lichtman2014big}.  We will
  focus on 3D ConvNets, which are relevant for both videos and 3D
  images.  2D ConvNets are regarded as a special case that is less
  computationally costly.

  In contemporary computer vision, researchers are most familiar with
  applying ConvNets to image-to-label transformations.  For example,
  in object recognition the input is an image of an object and the
  desired output is the class of the object.  However, ConvNets can
  also be used for image-to-image transformations.  In this context, a
  ConvNet acts like a more complex version of a typical filtering
  operation in image processing.  The ConvNet is applied to a window
  that slides over the input image.  For each location of the window,
  the ConvNet outputs a set of numbers, effectively producing a set of
  output images, each with the same resolution as the input image.

  For speeding up sliding window inference, a seemingly trivial detail
  about image borders turns out to be a bottleneck.  Inference is less
  efficient near the borders, and this inefficiency is proportionally
  smaller for a larger input image.  So efforts to maximize
  throughput tend to bump into the constraint of limited RAM.  (This
  is not the case for ConvNet training, where there is typically some
  range of input sizes that is optimal for training speed.)
  An apparently slower algorithm may end up having higher
  throughput if it can process a larger image within the constraint of
  the available RAM.

  \begin{figure}
    \begin{center}
      \includegraphics[width=0.79\columnwidth]{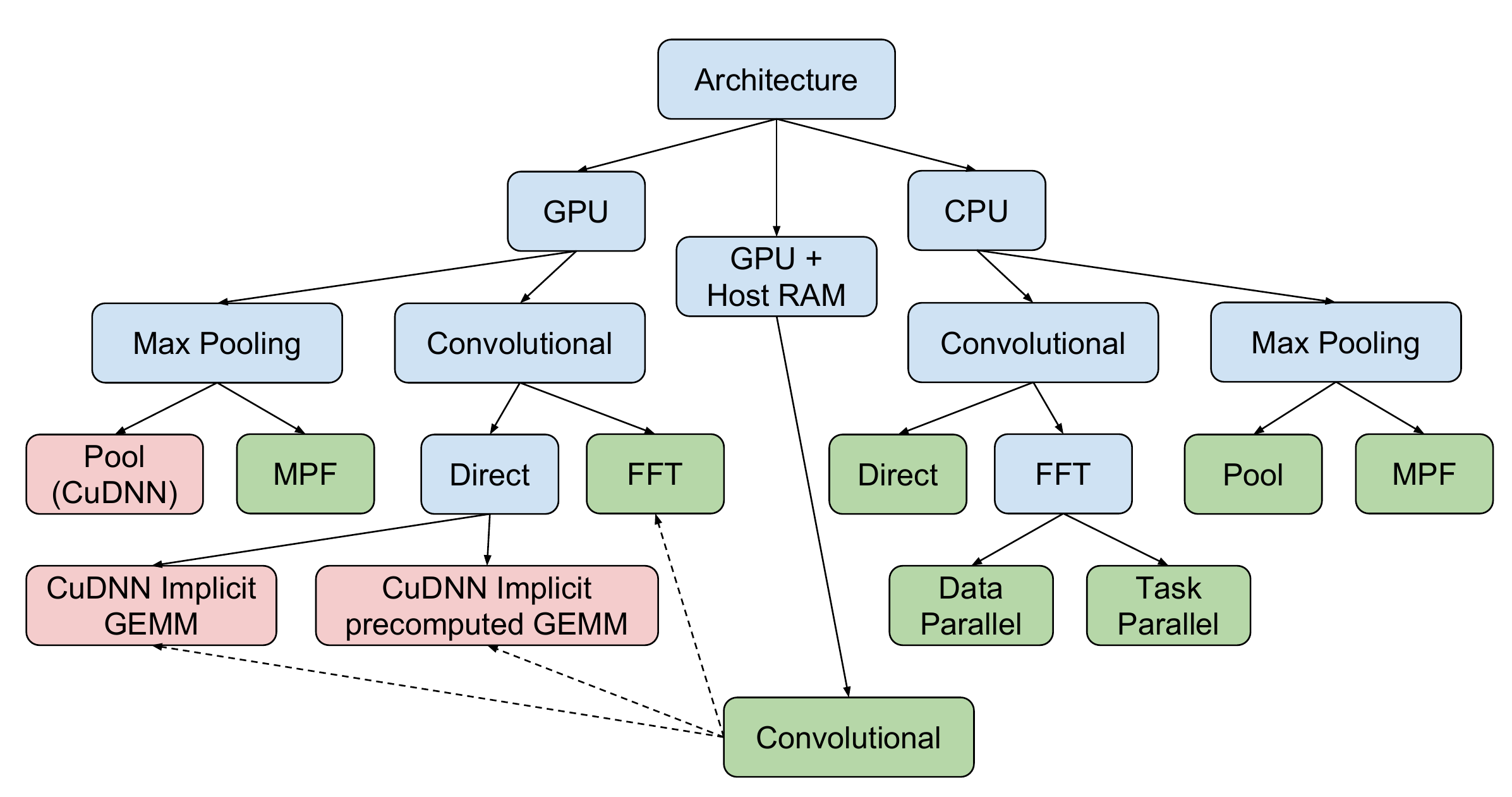}
    \end{center}
    \caption{Diagram of all layers primitives.  The red primitives are
      wrappers primitives provided by CuDNN.  The green primitives are
      the novel primitives introduced in this paper.}
    \label{fig:layers}
  \end{figure}

  We consider ConvNet architectures that contain both convolutional
  and pooling layers.  GPU implementations such as cuDNN provide a
  number of primitives for both kinds of layers.  A particular
  architecture is implemented by combining primitives.  In line with
  this approach, we introduce a number of new layer primitives for the
  CPU and GPU (Fig. \ref{fig:layers}).  These are designed to have low
  memory overhead, which should be important for high throughput as
  mentioned above.

  Our new convolutional primitives are either direct or FFT-based.
  For the latter, we introduce new implementations of pruned FFTs that
  are faster (for kernels on average $5\times$ for CPU and $10\times$
  for GPU) while using less memory.  Our FFT-based convolutional
  primitive for the GPU is designed to use much less memory than the
  algorithm proposed by~\cite{mathieu-iclr-14,vasilache2014fast} and
  implemented in {\tt fbfft}.  We provide two new FFT-based
  convolutional primitives for the CPU. The task parallel version is
  more efficient if the number of input images and the number of
  output images are both large, but at the cost of more memory
  overhead than the data parallel version.  The task parallel
  algorithm is designed to use less memory than the one previously
  proposed for ConvNet training~\cite{zlateski2015znn}.  To evaluate
  our novel primitives, we include comparisons with the direct
  convolutional primitives in cuDNN.  We also provide a new direct
  convolutional primitive for the CPU, but this turns out to be less
  useful than our FFT-based primitives in most circumstances.

  In empirical tests, we combine the primitives to maximize throughput
  for CPU-only and GPU-only algorithms. In some cases, our FFT-based
  GPU primitives outperform the cuDNN primitives by a large margin.
  The CPU-only algorithm may achieve higher throughput than the
  GPU-only algorithm because the CPU has fast access to more RAM, even
  if it is capable of fewer floating point operations (FLOPs) per
  second.

  To work around the limited onboard RAM of the GPU, we introduce a
  novel GPU + host RAM primitive.  We show that using this new
  primitive can lead to much higher throughput, in spite of the slower
  communication speed between GPU and host RAM.

  Finally, we study a CPU-GPU algorithm, in which the first layers of
  the ConvNet are computed by the CPU and the later layers are
  computed by the GPU.  The algorithm is pipelined to utilize the CPU
  and GPU efficiently.  This yields the highest throughput of all the
  algorithms, sometimes even greater than the sum of throughputs of
  GPU + host RAM and CPU-only.

  We also compare with other 3D sliding window max-pooling ConvNet
  implementations.  To the best of our knowledge, the only publicly
  available ones are the Caffe implementation
  of~\cite{tschopp2015efficient}, ZNN~\cite{zlateski2015znn}, and
  ELEKTRONN~\cite{ELEKTRONN2015}.  (The last is the only competitor
  that is specifically optimized for inference.) For comparison, we
  also include a baseline implementation that utilizes the simple
  pooling primitives of cuDNN to perform the naive algorithm of
  computing all subsamplings of the output.  For four representative
  ConvNet architectures running on a Titan X and/or a 4-way Intel Xeon
  E7 8890v3 machine with 256 GB RAM, our CPU-only, GPU-only, GPU +
  host RAM, and CPU-GPU implementations outperform all competitors.
  In particular, our CPU-GPU implementation wins by a factor of
  roughly $10\times$ or more in all cases.

  We focus on throughput (number of output voxels per unit time) as
  the performance metric.  A related metric is energy consumed per
  voxel.  For a given CPU or GPU, maximizing throughput is equivalent
  to minimizing energy consumption, assuming that the power
  consumption is roughly constant in time.  An interesting implication
  of our work is that in some situations the most economical way of
  increasing inference throughput may be to increase host RAM rather
  adding more GPUs or CPUs.



\section{Throughput of sliding window inference}
  Sliding window ConvNets were originally applied to detect and/or
  localize objects in a larger image~\cite{matan1991multi}, and this
  usage has been revived~\cite{sermanet2013overfeat}. Here each output
  image represents the probability that an object of a certain class
  is located at a voxel.  Sliding window ConvNets have also been
  applied to image segmentation, to produce an output image
  representing the probability that a voxel is a boundary between
  objects or not~\cite{jain2007supervised}.  And they have been
  applied to semantic segmentation, the problem of labeling each voxel
  in an input image by the class of the object to which it
  belongs~\cite{ning2005toward}.  In general, sliding window ConvNets
  are increasing in popularity as they are applied to more and more
  problems in computer vision.  However, transforming image-to-image
  is even more computationally costly than image-to-label, so the need
  for speeding up sliding window ConvNets is especially acute.

  In large scale sliding window inference, the input image is divided
  into smaller input patches. These are transformed by the ConvNet
  into output patches, which are recombined to generate the output
  image.  This divide-and-conquer approach is motivated by both time
  and space considerations.  The computation can be sped up by
  assigning the patches to multiple workers.  Also if the computation
  were not divided, it might not fit in the RAM available to a single
  worker.

  Each output patch is smaller than the input patch, because the
  sliding window is restricted to be entirely contained within the
  input image.  (This is analogous to a ``valid'' convolution in
  MATLAB.)  Therefore the input patches are chosen to overlap so that
  the output patches exactly cover the output image without
  overlap. (This is analogous to the overlap-save or overlap-scrap
  method for computing a single convolution described in signal
  processing textbooks.)

  We define the throughput of a single worker as the number of voxels
  in the output patch divided by the time required for the worker to
  process a single patch.  In this paper, a worker will be a single
  shared-memory machine, either CPU or GPU or combination of the two.
  Our goal is to maximize worker throughput.

  When computing ConvNet outputs for nearby locations of the sliding
  window, some of the arithmetic operations are identical.  For
  efficiency it is important to reuse the results of these operations.
  For ConvNets with convolutional layers only, this happens naturally
  as the output patch is the same resolution as the input patch.  If
  there are pooling layers, however, the output patch produced by a
  ConvNet is a subsampling of the output of a sliding window ConvNet.
  To obtain the entire output image, one must compute all subsamplings
  with different offsets separately, and then combine them.  This
  algorithm does not efficiently reuse computations for nearby output
  voxels.

  Here we instead provide pooling primitives that compute max-pooling
  fragments (MPF), a more efficient strategy for sliding window
  computations~\cite{giusti2013fast, masci2013fast}.  ELEKTRONN is the
  only other publicly available package with MPF support known to us
  (\url{http://elektronn.org}).  The MPF algorithm computes the same
  results as the approach known as ``dilated
  convolution''\cite{yu2015multi}, ``strided
  kernels''\cite{tschopp2015efficient}, ``max
  filtering''\cite{zlateski2015znn}, or ``filter
  rarefaction''\cite{long2015fully}.

  While MPF does efficiently reuse computation within a patch, the
  division into patches does itself incur a cost. Reuse cannot happen
  for nearby locations in different output patches, assuming the
  computations on the patches are done independently.  For example, it
  is highly inefficient to make the input patch the same size as the
  sliding window, which results in no reuse at all across different
  locations of the sliding window.  It is more efficient to increase
  the size of the input patch, to reduce the fraction of voxels near
  the border and thereby reduce inefficiency due to lack of reuse.

  Consequently memory overhead becomes important, if the goal is to
  maximize throughput.  An apparently faster algorithm may end up with
  inferior throughput if it cannot process a large input patch using
  the available RAM.  Later on, we will introduce a set of novel
  primitives for computing the layers of a ConvNet, which are designed
  to have low memory overhead as well as to be fast.  The primitives
  include convolutional and max-pooling layers.  (Our approach also
  can be extended to other kinds of pooling layers.)  As a prelude to
  the layer primitives, we first introduce lower-level FFT primitives
  which turn out to be important for efficient convolution.

\section{Pruned FFT}

  Improving the efficiency of convolution is crucial, because this
  computation is performed so many times in a ConvNet.  The advantages
  of FFT convolution have been shown for 2D ConvNets running on
  GPUs~\cite{mathieu-iclr-14,vasilache2014fast}, and 3D ConvNets
  running on CPUs~\cite{zlateski2015znn}.

  In FFT convolution the kernel and the image are zero-padded to a
  common size.  Since the kernel is typically much smaller than the
  image, the padded kernel consists mostly of zeros.  Ignoring the
  zeros is known as FFT pruning, and can provide speedup.  We propose
  and implement such an algorithm, for both the CPU and GPU. Our
  approach achieves an average of $5 \times$ speedup over the naive
  approach on the CPU and $10 \times$ speedup on the GPU.

  The speedup is expected to be large for a ConvNet, which typically
  contains many more kernel FFTs than image FFTs.  The speedup is more
  modest for a single convolution, which requires one padded kernel
  FFT, one image FFT, and one inverse FFT.

  While our pruned FFT algorithms give a substantial speedup,
  understanding them is not necessary for understanding the rest of
  our contributions.  The reader may prefer to skip to the next
  section on how pruned FFTs are used to compute the convolutional
  layers.

  \begin{figure}
    \begin{center}
      \includegraphics[width=0.49\columnwidth]{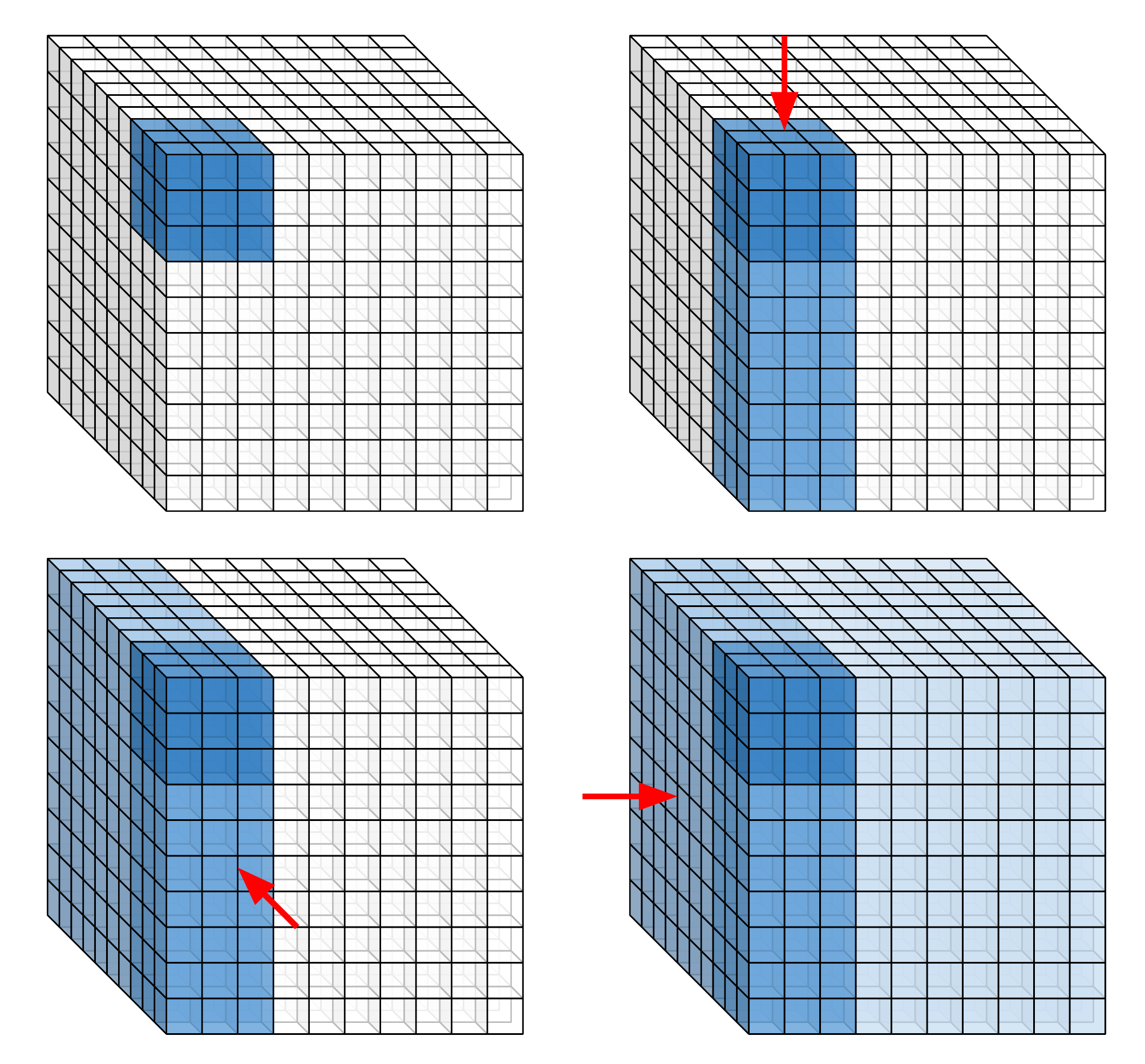}
    \end{center}
    \caption{Pruned FFTs. The dark blue voxels show the locations of
      the nonzero elements of the image after zero-padding. }
    \label{fig:pruned_ffts}
  \end{figure}

  \subsection{General algorithm}

  For 3D FFT-based convolution, the 3D images $x$ and $y$ are first
  zero-padded to the same size.  The inverse FFT of the point-wise
  product contains the result of the convolution.  The images $x$ and
  $y$ can be zero-padded to any size, as long as their size is equal.

  A 3D FFT is obtained by computing 1D FFTs along the three
  dimensions.  Some of these 1D FFTs are of an array with all elements
  equal to $0$.  These are unnecessary to compute as the FFT of
  an all zeros signal is all zeros.

  We can reduce the amount of computation by computing only necessary
  1D transforms.  When computing the FFT of a trainable kernel of size
  $k^3$ zero padded to size of $n^3$, instead of naively computing
  $n^2$ 1D FFTs along each dimension, which takes $C n^3 \log n^3$ we
  could first only do $k^2$ FFTs along one dimension, then $k \times
  x$ along then next, and finally $n^2$ along the last dimension, as
  shown on Fig.~\ref{fig:pruned_ffts}.  This approach reduces the
  computational cost from $C n^3 \log n^3$ to $C n\log n[k^2 + k \cdot
    n + n^2]$.  As most of the FFTs are performed on kernels, and as
  the kernel sizes are usually much smaller compared to the image
  sizes ($k \ll n$), we could reduce the computation cost by nearly
  two thirds.

  \subsection{CPU implementation}

  Suppose we would like to compute the FFT of an $x \times y \times z$
  image zero-padded to an $x' \times y' \times z'$ image.

  The $x \times y \times z$ image is zero-padded to $x' \times
  y \times z$.  This is easily implemented by doing a linear copy of
  the memory, and zero-padding the rest. We then perform $y \cdot z$
  1D real to complex FFT transforms along the $x$ direction.  The FFTs
  are performed out of place into a pre-allocated, and
  zero-initialized complex-valued $\floor{\frac{x'}{2}} + 1 \times
  y' \times z'$ image.  We then perform in-place 1D transforms along
  the $y$ direction, followed by the $z$ direction.

  The inverse FFT is computed by following the above steps in reverse
  order.  The 1D FFTs can either be done serially, or by $N$ workers
  in parallel (in a parallel for loop).

  This method induces a memory overhead of $x' \times y \times z$, a
  space required for zero--padding the image along $x$ direction in
  the first step of the algorithm.


  \subsection{GPU implementation}

  \label{sec:gpu-fft-impl}

  On the GPU, we always perform FFTs on $b$ 3D images simultaneously,
  in order to achieve high utilization of the many GPU threads.

  A set of $b$ 3D images can be represented as a 4D tensor.  We need
  to perform 1D FFTs along the three least significant dimensions.
  Our algorithm computes the 3D FFTs as a series of tensor
  transforms~\footnote{This is known as a \texttt{permute} function in
    MATLAB} and 1D FFTs along the least significant dimension of the
  tensor.

  When computing the FFT transforms of $b$ 3D images each of size $x
  \times y \times z$ padded to $x' \times y' \times z'$, the size of
  the 4D input tensor $I$ is $b \times x \times y \times z$.  First,
  1D in--place real to complex 1D transforms along the $z$ direction
  are performed.  We prepare the input by extending the 4D tensor
  along the $z$ direction to fit the result.  The transform will need
  to contain $z'' = z' / 2 + 1$ complex numbers, and we need twice as
  many reals.  A 4D tensor $I^1$ of size $b \times x \times y \times
  2z'')$ is first initialized to zero, and appropriate elements from
  the input are filled in (elements $I_{i,j,k,l}$ get mapped to
  $I^1_{i,j,k,l}$ while the rest of elements of $I^1$ are set to
  zero).

  A batch of $b$ in--place real to complex 1D transforms are then
  performed.  The result represents a 4D complex tensor
  $\widetilde{I}^1$ of size $b \times x \times y \times z''$.  Note
  that all the 1D transforms are done on contiguous memory chunks
  (along the least significant dimension).

  In the next step we perform in-place complex to complex transforms
  along the $y$ direction.  To do this the elements of
  $\widetilde{I}^1$ are permuted into another 4D tensor
  $\widetilde{I}^2$ of size $b \times x \times z'' \times y'$, such
  that the element $\widetilde{I}^1_{i,j,k,l}$ gets mapped to
  $\widetilde{I}^2_{i,j,l,k}$ and the rest of $\widetilde{I}^2$ is
  zero--filled.  We then perform in-place complex to complex
  transforms along the least significant dimension of
  $\widetilde{I}^2$.

  In the final step, we perform the transform along the $x$ direction.
  We permute $\widetilde{I}^2$ into a new 4D complex tensor
  $\widetilde{I}^3$ of size $b \times z'' \times y' \times x'$.  An
  element $\widetilde{I}^2_{i,j,k,l}$ is mapped to
  $\widetilde{I}^3_{i,k,l,j}$.  Complex to complex in-place transforms
  are performed along the least significant dimension of
  $\widetilde{I}^3$.

  As we only perform point-wise multiplications of transformed images,
  or take the inverse transforms of them, we can just keep the result
  in this representation -- not waste computation doing extra
  permuting.

  The inverse transform can be performed taking the same steps in
  reverse order.

  \subsection{Implementation details}

  Our approach uses $3$rd party libraries to perform each batch of 1D
  FFTs.  Depending on the library implementation, the size to which we
  pad the 3D image can greatly influence the computational complexity.

  On the CPU we use either {\tt fftw} or {\tt Intel MKL}, and pad the
  images (and kernels) to sizes that can be written in the form of
  $2^a3^b5^c7^d11^e13^f$.  When {\tt Intel MKL} is used any such size
  is allowed, however, when {\tt fftw} is used we only allow sizes for
  which $e+f$ is either $0$ or $1$~\cite{frigo1999fftw,frigo1998fftw}.
  On the GPU we use cuFFT~\cite{nvidia2010cufft}, which has optimized
  algorithms only for sizes of the form $2^a3^b5^c7^d$.

  4D tensor permuting requires a lot of indexing calculation, which
  can involve a lot expensive division and modulus operations.
  Sometimes these operations are more expensive than the actual 1D FFT
  transforms performed.  We improve the performances by only using
  multiplications by a pre--computed magic numbers and shift
  operations as described in ~\cite{warren2013hacker}.  Image
  reshaping is easily implemented using the Thrust CUDA
  library~\cite{bell2011thrust}.

  We limit the large cuFFT memory overhead for computing batches of 1D
  transforms by splitting the batch computation into sub--batches of
  1D transforms.  We make sure that the sub--batch size is still large
  enough to utilize all the computational power, but limit the size so
  that we limit the memory overhead.

  The memory overhead of the algorithm is due to the fact that we do
  out-of-place permuting of the 4D tensor, and requires space for $b
  \cdot x \cdot y' \cdot z''$ complex numbers.  This, however, will
  not increase the memory overhead of our algorithm for convolutional
  layers on the GPU, as it already needs a temporary tensor of size $b
  \cdot x' \cdot y' \cdot z''$ for other purposes, which we can use as
  the scratch space for computing the FFT transforms.

  Additional, relatively small, overhead comes from the memory
  required by cuFFT to perform a batch of 1D transforms.  By dividing
  the batch into sub--batches we essentially limit this overhead to a
  pre--defined constant amount of memory.

\section{Convolutional layers}

  We begin with the primitives for the convolutional layers, which are
  the most computationally intensive.

  The input to a convolutional layer is a tuple of $f$ images, and the
  output a tuple of $f'$ images.  We want to process a batch of $S$
  inputs to yield a batch of $S$ outputs, via
  \[
  O_{s,j} = \sum_{i=1}^f w_{ji}\ast I_{s,i}
  \]
  for $1 \le s \le S$ and $1 \le j \le f'$.  Here $I_{s,i}$ is the
  $i^{th}$ image of the $s^{th}$ input in the batch, and $O_{s,j}$ is
  the $j^{th}$ image of the $s^{th}$ output in the batch, and $w_{ji}$
  is the kernel from the $i^{th}$ image in an input tuple to the
  $j^{th}$ image in an output tuple.

  We will assume 3D images and kernels.  If $I_{s,i}$ has size
  $\vec{n} = \angled{n_x, n_y, n_z}$ and $w_{ji}$ has size $\vec{k} =
  \angled{k_x,k_y,k_z}$, then we can regard $I$ as a 5D tensor of size
  $S \times f \times n_x \times n_y \times n_z$, $w$ as a 5D tensor of
  size $f' \times f \times k_x \times k_y \times k_z$, and $O$ as a 5D
  tensor of size $S \times f' \times n_x' \times n_y' \times n_z'$,
  where $\vec{n}' = \vec{n} - \vec{k} + \vec{1}$.

  We will refer to the sizes of the 5D tensors $I$ and $O$ as input
  and output shape, respectively.  The relationship between input
  shape and output shape depends on kernel size as in
  Table~\ref{table:layers_complexity}.

  \begin{table*}[t]
    {\footnotesize
    \centering
    \begin{tabular}{l|lll}
      \toprule
      Layer   & Input shape    & Output shape     & FLOPS \\
      \midrule
      Convolutional -- Direct &
      $S \times f \times n^3$ &
      $S \times f' \times [n-k]^3$ &
      $S \cdot f' \cdot f \cdot n^3 \cdot k^3$ \\
      Convolutional -- FFT--based &
      $S \times f \times n^3$ &
      $S \times f' \times [n-k]^3$ &
      $S \cdot 3Cn^3 \log n[f' + f] + 4Sf' \cdot f \cdot n^3 + f \cdot f' \cdot Cn \log n[k^2 + k \cdot n + n^2]$ \\
      Max Pooling &
      $S \times f \times n^3$ &
      $S \times f \times [n/p]^3$ &
      $S \cdot f \cdot n^3$ \\
      Max Fragment Pooling &
      $S \times f \times n^3$ &
      $[S \cdot p^3] \times f \times (n/p)^3$ &
      $S \cdot f \cdot n^3 \cdot p^3$ \\
      \bottomrule
    \end{tabular}
    \caption{Relation between input and output shapes for
      convolutional and pooling layers, along with computational
      complexities. Input shape is for a batch of $S$ inputs, each of
      which is an $f$-tuple of 3D images with size $n^3$, and output
      shape is analogous. The 3D kernel has size $k^3$. The pooling
      window has size $p^3$, and the constant $C$ for the FFT
      complexity depends on the FFT implementation.}
    \label{table:layers_complexity}
    }
  \end{table*}

\subsection{CPU algorithms}

  We propose three parallel algorithms for the convolutional layer
  that are suited for multi-core CPUs.  The first algorithm performs
  direct convolution, whereas the other two use FFT based
  convolutions.

\subsubsection{Direct convolution algorithm}

  \begin{algorithm}
    {\footnotesize
      \begin{codebox}
        \Procname{$\proc{Convolutional-Forward-FFT-CPU1}(I,w,S,f,f',\vec{n},\vec{k})$}
        \li $\vec{n}' = \vec{n} - \vec{k} + \vec{1}$
        \li $O \gets \proc{5D-Real-Tensor}(S,f',n'_x,n'_y,n'_z)$
        \li \kw{parallel for} $i \gets 0 \To S-1$
        \li   \Do \kw{parallel for} $j \gets 0 \To f'-1$
        \li     \Do \For $k \gets 0 \To f-1$
        \li     \Do $O_{i,j} \gets O_{i,j} + \proc{Convolve}(I_{i,k},w_{j,k})$
        \End \End \End
        \li $\proc{Free-Memory}(I)$
        \li \Return $O$
      \end{codebox}
    \caption{Multi-core algorithm for a convolutional layer using direct
      convolution.}
    \label{alg:cpu_direct}
    }
  \end{algorithm}

  The computation is parallelized by two $\kw{parallel for}$ loops
  such that each image of each output in the batch is computed in
  parallel on a different working thread (see
  Algorithm~\ref{alg:cpu_direct}).  The $\kw{parallel for}$ loops are
  implemented using Intel thread building blocks such that the work is
  evenly divided over the available cores.

  We provide one implementation using naive convolution and the other
  using Intel MKL.  The latter is $2\times$ faster on average, but
  requires extra memory for a temporary image where a result of
  convolution is stored before accumulating it to the output image.
  The memory overhead of both implementations is given in
  Table~\ref{table:memory_requirements}.

  \begin{table}
    {\footnotesize
    \centering
    \begin{tabular}{l >{$}l<{$}}
      \toprule
      CPU algorithm & \text{Memory required} \\
      \midrule
      Direct (naive) &
      S \cdot f \cdot n + S \cdot f' \cdot n'\\
      Direct (MKL) &
      S \cdot f \cdot n + S \cdot f' \cdot n' + T \cdot n' \\
      FFT algorithm 1 &
      \max
      \begin{cases}
        S \cdot f \cdot (n + \widetilde{n}) \\
        S \cdot f' \cdot n' + (S \cdot f + 1) \cdot \widetilde{n}
      \end{cases} \\
      FFT algorithm 2 &
      \max
      \begin{cases}
        S \cdot f \cdot (n + \widetilde{n}) \\
        S \cdot (f + f') \cdot \widetilde{n} + T \cdot \widetilde{n} \\
        S \cdot f' \cdot (n' + \widetilde{n})
      \end{cases} \\
      \bottomrule
      \toprule
      GPU algorithm & \text{Memory required} \\
      \midrule
      cuDNN (default) &
      S \cdot f \cdot n + S \cdot f' \cdot n' \\
      cuDNN (precomp) &
      2S \cdot f \cdot n + S \cdot f' \cdot n' \\
      FFT &
      K + \max
      \begin{cases}
        S \cdot f \cdot (n + \widetilde{n}) + f \cdot \widetilde{n} \\
        S \cdot (f + f') \cdot \widetilde{n} + 2f \cdot \widetilde{n} \\
        S \cdot f' \cdot (n' + \widetilde{n}) + f' \cdot \widetilde{n}
      \end{cases} \\
      \bottomrule
    \end{tabular}

    \caption{Memory required by different implementations.  $S$ is the
      batch size, $f$ and $f'$ represent the number of input/output
      images of the layer.  $n$ and $n'$ represent the number of
      pixels in each input/output image, and $\widetilde{n}$
      represents the number of elements in the transformed image.  $K$
      is pre--defined constant amount of memory allocated for cuFFT,
      and $T$ is the number of available cores for the CPU
      algorithms.}
    \label{table:memory_requirements}
    }
  \end{table}

\subsubsection{Data parallel FFT-based algorithm}

  The computationally intensive operations are individually
  parallelized (see Algorithm~\ref{alg:cpu_fft_alg1}).  More
  specifically each FFT and inverse FFT transform is done in parallel
  as explained in the previous section.  The $\proc{Parallel-MAD}$
  function computes a series of multiply-add operations of complex
  numbers in parallel by dividing the range into roughly equal
  sub-ranges, each of which is executed on a single core.

  \begin{algorithm}
    {\footnotesize
      \begin{codebox}
        \Procname{$\proc{Convolutional-Forward-FFT-CPU1}(I,w,S,f,f',\vec{n},\vec{k})$}
        \li $\vec{n}' = \vec{n} - \vec{k} + \vec{1}$
        \li $\vec{\widetilde{n}} = \proc{FFT-Optimal-Size}(\vec{n})$
        \li $\widetilde{I} \gets \proc{5D-Complex-Tensor}(S,f,\floor{\widetilde{n}_x/2}+1,\widetilde{n}_y,\widetilde{n}_z)$
        \li \For $i \gets 0 \To S-1$
        \li   \Do \For $j \gets 0 \To f-1$
        \li     \Do $\widetilde{I}_{i,j} \gets \proc{Parallel-FFT}(I_{i,j})$
        \End \End
        \li $\proc{Free-Memory}(I)$
        \li $O \gets \proc{5D-Real-Tensor}(S,f',n'_x,n'_y,n'_z)$
        \li $\widetilde{O} \gets \proc{4D-Complex-Tensor}(S,\floor{\widetilde{n}_x/2}+1,\widetilde{n}_y,\widetilde{n}_z)$
        \li $\widetilde{w} \gets \proc{3D-Complex-Tensor}(\floor{\widetilde{n}_x/2}+1,\widetilde{n}_y,\widetilde{n}_z)$
        \li \For $i \gets 0 \To f'-1$
        \li   \Do \For $j \gets 0 \To f-1$
        \li     \Do $\widetilde{w} = \proc{Parallel-FFT}(w_{i,j})$
        \li         \For $k \gets 0 \To S-1$
        \li           \Do $\proc{Parallel-MAD}(\widetilde{I}_{k,j}, \widetilde{w}, \widetilde{O}_{k})$
        \End \End
        \li   \For $k \gets 0 \To S-1$
        \li      \Do $O_{k,i} \gets \proc{Parallel-Inverse-FFT}(\widetilde{O}_{k})$
        \End \End
        \li $\proc{Free-Memory}(\widetilde{I})$
        \li $\proc{Free-Memory}(\widetilde{O})$
        \li $\proc{Free-Memory}(\widetilde{w})$
        \li \Return $O$
      \end{codebox}
    }

    \caption{Multi-core algorithm for a convolutional layer}
    \label{alg:cpu_fft_alg1}
  \end{algorithm}

  The memory requirement of the algorithm equals to the maximal amount
  of memory required by the algorithm at any single point of time
  during the execution, and is given in
  Table~\ref{table:memory_requirements}.

\subsubsection{Task parallel FFT-based algorithm}

  The main quantities of the task parallel algorithm are: (1) breaking
  up the computation required by the convolutional layer into tasks
  that operate on independent chunks of memory, (2) creating a task
  dependency graph, and (3) scheduling the tasks for execution.

  There are five different task types:

  {\color{zblack}}

  \begin{itemize}
    \item {\color{zred}\bf Input image transform} task computes the
      forward FFT transform of a single input image.
    \item {\color{zblue}\bf Kernel transform} task computes the forward
      FFT transform of a single kernel.
    \item {\color{zgreen}\bf Multiply-add} task computes the
      point-wise product of an input image and a kernel FFT
      accumulating the result to an appropriate image transform.
    \item {\color{zpurple}\bf Output image transform} task computes
      the inverse FFT of the appropriate accumulated image transform.
      This task is also responsible for adding the bias and applying
      the transfer function.
    \item {\color{zyellow}\bf Synchronization} tasks, beside serving
      as synchronization points are the only tasks responsible (and
      only ones allowed) to allocate and/or deallocate memory.
  \end{itemize}

  {\color{zblack}}

  \begin{figure}
    \begin{center}
      \includegraphics[width=0.85\columnwidth]{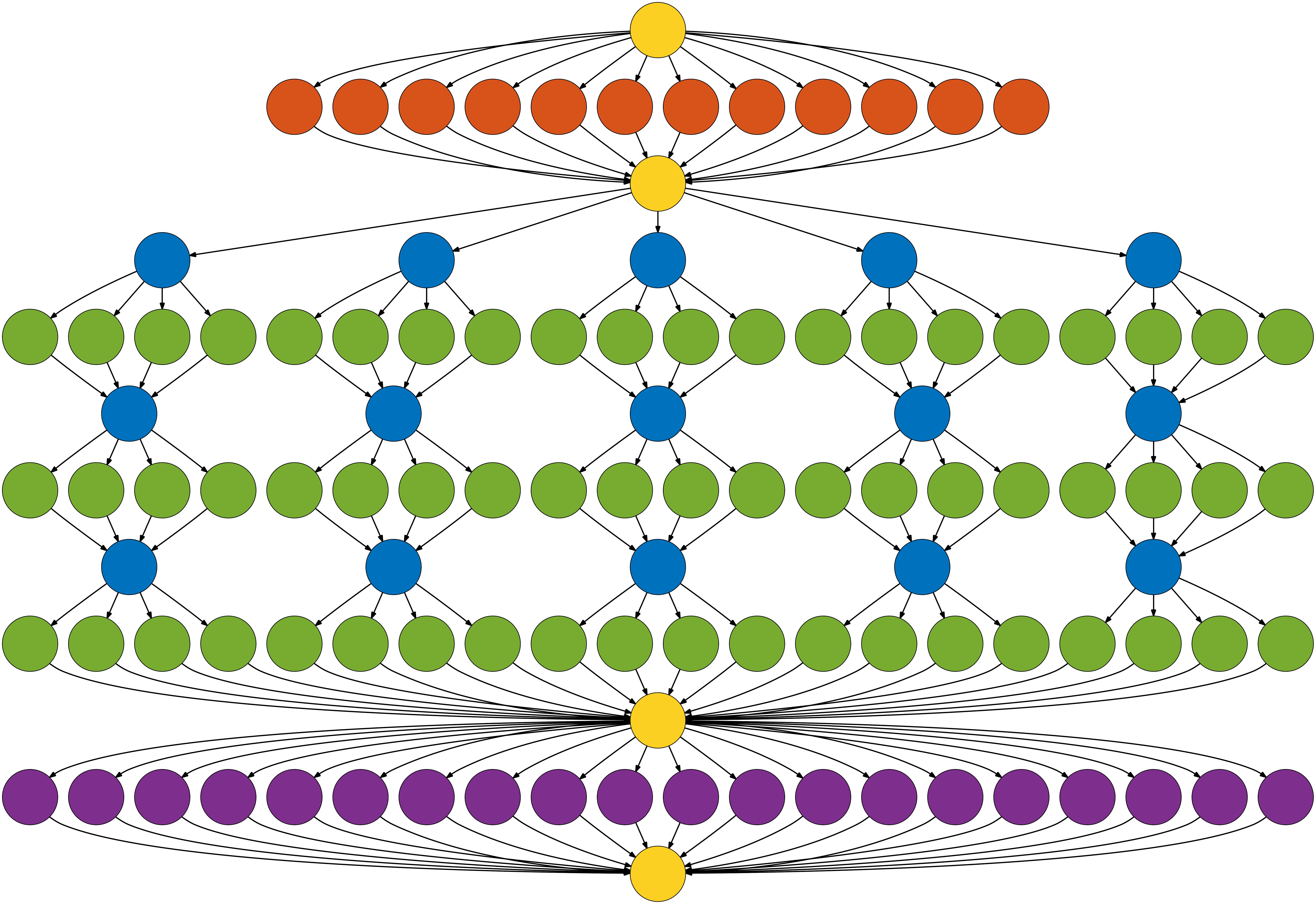}
    \end{center}
    \caption{Task dependency diagram of a task--based convolutional
      layer.}
    \label{fig:task_deps}
  \end{figure}

  The task dependency graph of all the tasks required for computing
  the output images of a convolutional layer with four input and five
  output images for a batch size of four is shown on
  Fig.~\ref{fig:task_deps}.  The tasks are created and queued when all
  their dependencies have been satisfied.  There are four {\bf
    synchronization} tasks effectively dividing the computation into
  three stages.  The layout of the {\bf kernel transform} tasks forms
  a grid, with the number of columns equal to the number of output
  images, and the number of rows equal to the number of input images.
  The task of the $i^{th}$ column and $j^{th}$ row computes the
  transform of the kernel $w_{i,j}$.  Furthermore, each such task has
  $S$ dependent {\bf multiply-add} tasks, where $S$ is the batch size
  of the input (equal to four in Fig.~\ref{fig:task_deps}).  The
  $k^{th}$ dependent {\bf multiply-add} task of a {\bf kernel
    transform} task in column $i$ and row $j$ accumulates the product
  of the transforms of the $j^{th}$ input image of the $k^{th}$ batch
  and the filter $w_{i,j}$ to the transform of the $i^{th}$ output
  image of the $k^{th}$ batch.

  The tasks are executed by $N$ worker threads, where $N$ equals the
  number of available cores (or virtual cores, when hyper--threading
  is enabled).  Each worker thread is {\bf pinned} to a single core.
  This means that there is 1--1 relation between the workers and the
  available cores -- each worker is allowed to run only on a specific
  hardware core, as described in~\cite{jeffers2015high}.  For this
  reason, we will use ``worker thread'' and ``core'' interchangeably.

  The first {\bf synchronization} task allocates memory for the FFT
  transforms of the input images.  The number of dependent {\bf input
    image transform} tasks equals the number of input images times the
  batch size $S$.  They are then executed by the $N$ worker threads,
  such that each worker picks up an arbitrary task and executes it.

  The last thread to complete the execution of an {\bf input image
    transform} task immediately executes the second {\bf
    synchronization} task.  This tasks deallocates the memory holding
  the input images, as their values are no longer required. It then
  allocates memory for the transforms of the output images.  At this
  point $M$ threads are chosen as \emph{primary} threads, where $M$ is
  the maximum of $N$ -- total number of threads, and the number of
  output images.  The \emph{primary} threads are chosen so that they
  are evenly distributed over multiple physical chips.  Each
  \emph{primary} thread is given a temporary buffer equal to the size
  required to fit the transform of a padded kernel for that layer.

  The {\bf kernel transform} tasks and {\bf multiply-add} tasks are
  then scheduled for execution based on their distance to the sink
  node of the task dependency graph, such that the more distant nodes
  get scheduled first.  The scheduling has two additional constraints:
  (1) the {\bf kernel transform} tasks can only be executed by a
  \emph{primary} thread, and (2) its dependent {\bf multiply-add}
  tasks can only be executed by worker threads that are pinned to the
  cores on the same physical chip.

  This strategy is chosen over the popular alternative approach to
  task scheduling based on work stealing~\cite{reinders2007intel,
    willhalm2008putting} because it divides the work more evenly over
  multi--chip machines and further increase cache locality.  On a
  4-way machine it yields more deterministic results -- very little
  variance in run-time and average of 20\% speed improvement over the
  alternative.

  The last {\bf multiply-add} task to complete executes the third {\bf
    synchronization} task.  This task deallocates the memory buffers
  given to the primary threads as well as the memory used to store the
  transforms of the input images.  It also allocates the memory to
  store the final result -- the output images.

  The number of {\bf output image transform} tasks equals the number
  of output images times the batch size.  The tasks are executed by
  all $N$ workers, such that each worker picks up an arbitrary task
  and executes it.  The last {\bf output image transform} task to
  finish also executes the final {\bf synchronization} task, which
  frees the memory required for the output image transforms.

  The task parallel algorithm requires that both $f \cdot S$ and $f'
  \cdot S$ be large enough (at least the same size as the number of
  available cores) in order to efficiently utilize all the cores.  In
  such cases they can be much more efficient than the data parallel
  algorithm, as the tasks operate on independent data (thus reducing
  false--sharing~\footnote{A case where one CPU's cache is invalidated
    because multiple CPUs operate on adjacent memory}).  It further
  improves the performances on multi--chip systems, by minimizing the
  data sharing between cores on different chips. On a 4--way Intel
  Xeon E7--8890 v3 machine the task parallel algorithm is $10 \times$
  faster than the data parallel one (for large enough $f' \cdot S$ and
  $f \cdot S$).

  As shown later, this is the implementation that is optimal in most
  cases, even for very small kernel sizes; the only exception is the
  first layer of the network for which both $f = 1$ and $S = 1$.

  The memory required by the task parallel algorithm can be higher
  then the one of the data parallel algorithm, when many cores are
  available.  The exact memory required equals to the maximal memory
  required by each of the 3 stages, and is given in
  Table~\ref{table:memory_requirements}.

\subsection{GPU implementations}

  For the GPU, we propose three different algorithms.  Two of them use
  cuDNN's 3D primitives that are based on implicit matrix--matrix
  multiplication.  The third FFT--based implementation is based on
  our, previously described, algorithm for pruned FFTs.

\subsubsection{Direct convolution using cuDNN}

  The two algorithms using the direct convolution are implemented
  using cuDNN.  CuDNN performs 3D convolution as an implicit
  matrix--matrix multiplication, meaning that the matrices are not
  actually created.  The first algorithm improves the speed by
  pre--computing a set of indices, and thus require additional
  workspace memory.  The second algorithm, which we find 3-5$\times$
  slower does not require any extra memory.

\subsubsection{FFT based algorithm}

  FFT-based convolutional layer is based on the GPU implementation of
  the pruned FFT algorithm described in
  Section~\ref{sec:gpu-fft-impl}.

  \begin{algorithm}
    {\footnotesize
    \begin{codebox}
      \Procname{$\proc{Convolutional-Forward-FFT-GPU}(I,w,S,f,f',\vec{n},\vec{k})$}
      \li $\vec{n}' = \vec{n} - \vec{k} + \vec{1}$
      \li $\widetilde{I} \gets \proc{5D-Complex-Tensor}(S,f,\floor{n_x/2}+1,n_y,n_z)$
      \li $\widetilde{s} \gets \proc{5D-Complex-Tensor}(f,\floor{n_x/2}+1,n_y,n_z)$
      \li \For $i \gets 0 \To S-1$
      \li   \Do $\widetilde{I}_{i} \gets \proc{GPU-Parallel-FFT}(I_{i},\widetilde{s})$
      \End
      \li $\proc{Free-Memory}(I)$
      \li $\widetilde{O} \gets \proc{5D-Complex-Tensor}(S,f',\floor{n_x/2}+1,n_y,n_z)$
      \li \For $i \gets 0 \To f'-1$
      \li    \Do $\widetilde{w}_{i} = \proc{Parallel-FFT}(w_{i},\widetilde{s})$
      \li        \For $j \gets 0 \To S-1$
      \li           \Do $s \gets \proc{Parallel-Mult}(\widetilde{w}_{i}, \widetilde{I}_{j})$
      \li               $\widetilde{O}_{j,i} \gets \proc{Parallel-Accumulate}(s)$
      \End \End
      \li $\proc{Free-Memory}(\widetilde{I})$
      \li $\proc{Free-Memory}(\widetilde{s})$
      \li $O \gets \proc{5D-Real-Tensor}(S,f',n'_x,n'_y,n'_z)$
      \li $\widetilde{s} \gets \proc{5D-Complex-Tensor}(f',\floor{n_x/2}+1,n_y,n_z)$
      \li \For $i \gets 0 \To S-1$
      \li   \Do $O_{i} \gets \proc{GPU-Parallel-Inverse-FFT}(\widetilde{O}_{i})$
      \End
      \li $\proc{Free-Memory}(\widetilde{O})$
      \li $\proc{Free-Memory}(\widetilde{s})$
      \li \Return $O$
    \end{codebox}
    }

    \caption{FFT based convolutional layer algorithm for the GPU.}
    \label{alg:gpu_alg}
  \end{algorithm}

  The algorithm, given in Algorithm~\ref{alg:gpu_alg} resembles the
  task based CPU algorithm in the sense that it consists of three
  stages with memory allocated/deallocated between the stages.  The
  lines 2 and 3 allocate memory required for the input image
  transforms and the scratch space required by
  $\proc{GPU-Parallel-FFT}$ procedure (explained in
  Section~\ref{sec:gpu-fft-impl}).  The first stage (lines 4 and 5)
  computes the transforms of all the input images by performing $f$ 3D
  FFTs in parallel.  The memory used by the input images is then
  released, and memory for storing the FFTs of the output images is
  allocated (lines 6 and 7).

  In the second stage (lines 8--12) we loop over the $f'$ output
  images.  For each output image we compute the transform
  of the $f$ relevant kernels (ones connecting each of the input
  images and the current output image).  We then loop
  over the inputs in the batch, and for each batch we compute the
  point--wise product of the relevant input image transforms with the
  relevant kernel transforms, producing $f$ complex valued images.
  The values of the $f$ images are then accumulated to a single image,
  -- the transform of the appropriate output image.  Note how we can
  re--use the scratch space $s$ (used for $\proc{GPU-Parallel-FFT}$)
  to store the point--wise product of $f$ transformed images.

  The memory used by the input image transforms, and the scratch space
  is then released.  We then allocate memory for the output images as
  well as new scratch space of different size, required for computing
  $f'$ inverse FFT transforms at once (lines 13--16).

  In the final stage (lines 17 and 18) we compute the output images by
  looping over the batches and computing $f'$ inverse FFTs in
  parallel.  Finally we free the memory of the output transforms and
  the scratch space.

  The memory required by the algorithm is equal to the maximum of
  memory required in each of the three stages
  (Table~\ref{table:memory_requirements}).

\section{Max-pooling and max-pooling fragments}

  \emph{Max pooling} of an image of size $\vec{n}$ with the window
  size of $\vec{p} = \angled{p_x,p_y,p_z}$ divides an image into
  blocks of size $\vec{p}$.  The maximum value is computed for each
  block, yielding an image of size $\angled{n_x/p_x,n_y/p_y,n_z/p_z}$.
  The input image size $\vec{n}$ is restriceted such that $n_x$, $n_y$
  and $n_z$ are divisible by $p_x$, $p_y$ and $p_z$ respectively.

  On the CPU, we implement the max-pooling layer so that the
  max-pooling of each image is performed in parallel (e.g. by using
  parallel for loop).  For the GPU we use the cuDNN primitives for
  max-pooling.

  When the input image has the same size as the ConvNet field of view,
  the output image consists of a single voxel.

  \emph{Max pooling fragmentation} of an image of size $\vec{n}$ with the
  window size of $\vec{p}$ produces $p_x \times p_y \times p_z$ output
  images (fragments) by performing multiple \emph{max pooling}
  operations on the image at offsets $(x,y,z)$, where $0 \le x < p_x$,
  $0 \le y < p_y$, and $0 \le z < p_z$.  When the image has size such
  that $\vec{n} + \vec{1}$ is divisible by $\vec{p}$, the sizes of all
  produced fragments will equal
  $\angled{\floor{n_x/p_x},\floor{n_y/p_y},\floor{n_z/p_z}}$.

  It is important to note that max-pooling fragmentation increases the
  batch size for subsequent layers.  For an MPF layer, the number of
  output images is equal to the number of input images times the
  number of fragments $p_x \times p_y \times p_z$.  The increase in
  the batch size has an impact on the parallelization of subsequent
  layers.  Simple max-pooling does not change the batch size.

  Our CPU implementation loops over all the $f$ input images of each
  of $S$ inputs in a parallel for loop, and performs the max-pooling
  operation at each offset.

  In the GPU implementation, for each valid offset $(x,y,z)$ we invoke
  the cuDNN max-pooling primitive to compute the max-pooling of all
  input images at that offset.

  \begin{table}
    {\footnotesize
    \centering
    \begin{tabular}{cccccc}
      \toprule
      Layer & F--maps & n337    & n537  &  n726  &  n926 \\
      \midrule
      $1$ & 80 &  Conv $2^3$  & Conv $4^3$  & Conv $6^3$  & Conv $8^3$ \\
      $2$ & 80 &  Pool $2^3$  & Pool $2^3$  & Pool $2^3$  & Pool $2^3$ \\
      $3$ & 80 &  Conv $3^3$  & Conv $5^3$  & Conv $7^3$  & Conv $9^3$ \\
      $4$ & 80 &  Pool $2^3$  & Pool $2^3$  & Pool $2^3$  & Pool $2^3$ \\
      $5$ & 80 &  Conv $3^3$  & Conv $5^3$  & Conv $7^3$  & Conv $9^3$ \\
      $6$ & 80 &  Pool $2^3$  & Pool $2^3$  & Conv $7^3$  & Pool $9^3$ \\
      $7$ & 80 &  Conv $3^3$  & Conv $5^3$  & Conv $7^3$  & Conv $9^3$ \\
      $8$ & 80 &  Conv $3^3$  & Conv $5^3$  & Conv $7^3$  & Conv $9^3$ \\
      $9$ & 80 & Conv $3^3$  & Conv $5^3$  & & \\
      $10$ & 3 & Conv $3^3$  & Conv $5^3$  & & \\
      \bottomrule
    \end{tabular}
    \caption{ConvNet architectures of the benchmarked networks.}
    \label{table:benchmarked_networks}
    }
  \end{table}

\section{GPU-only or CPU-only inference}
  By stringing together the CPU (or GPU) layer primitives defined
  above, we can now construct CPU-only (or GPU-only) algorithms for
  ConvNet inference.  For each layer, we have a choice between several
  primitives.  Each \emph{max--pooling} layer can be replaced by
  a \emph{MPF} layer.  The size of the input patch and the number of
  inputs in the batch should be chosen.  These parameters and the
  primitives should be chosen to maximize throughput.  Below we
  describe some theoretical considerations and empirical data about
  the optimal choice.

\subsection{Maximizing throughput}
  The result of a network applied to an input $I$ is obtained by
  sequentially applying each primitive.  The input to the first layer's
  primitive is $I$, and the input to every subsequent primitive will
  be the output of the previous layer's primitive.  The output of the
  last layer will be our desired result $I'$.  Note that if \emph{MPF}
  layers are used, the most significant dimension of the output can
  increase.  For an input shape of $(S,f,x,y,z)$ the output shape will
  be of the form $(\alpha S,f',x',y',z')$.  Where $\alpha$ value is
  depends on the amount of \emph{MPF} layers used and their pooling
  window sizes.  This output represents $S$ sets, each having $\alpha$
  fragments which should be recombined to obtain the sliding--window
  result.~\cite{giusti2013fast,masci2013fast}.

  The throughput of the network is defined as:
  \[
  \frac{\proc{Size}(I')}{\sum_{1 \le i \le L}
    \proc{Time}(\id{Primitive}_i,I_i)}
  \]
  Where $I_i$ is the input of the $i^{th}$ layer's primitive.  The
  output shape will depend on the shape of the input $I$ and the
  primitives chosen for each \emph{max--pooling} layer.  As the input
  shapes to each layer need to have integral sizes, not every
  combination of layer primitives and input shapes are allowed (see
  Table~\ref{table:layers_complexity}).  Additional constraint is that
  the memory required for $i^{th}$ primitive to process input $I_i$
  has to be smaller than the memory available to the system (either
  the CPU or the GPU).

  In general case, the highest throughput network implementation can
  be found using an exhaustive search:

  \begin{enumerate}
    \item Loop over all possibilities for the \emph{max--pooling}
      layers.  This will introduce constraints on the allowable input
      shapes.
    \item Loop over all allowed input shapes.
    \item Determine the best implementation for each convolutional
      layer.
  \end{enumerate}

  This is possible because for fixed choice of \emph{max--pooling} or
  \emph{MPF} of each pooling layer, and fixed input shape, the time
  and space required for each convolutional layer is uniquely
  determined.  We pick the fastest one that satisfies the memory
  constrain.

  In the empirical measurements below, it will turn out that for our
  networks, the highest throughput is obtained when all
  the \emph{max--pooling} layers are replaced with \emph{MPF} layers,
  and when the input batch size is one ($S = 1$).  Additionally,
  higher throughput is achieved for larger input sizes.

  The fact that \emph{MPF} layers outperform \emph{max--pooling
    layers} is not surprising as it has been shown that
    using \emph{MPF} layers reduces amount of operations required for
    computing a single output
    pixel~\cite{giusti2013fast,masci2013fast}.

  \begin{figure}
    \centering
    \subfloat[]{\protect\includegraphics[trim= 0mm 0mm 9mm 1mm, clip, width=0.5\columnwidth]
      {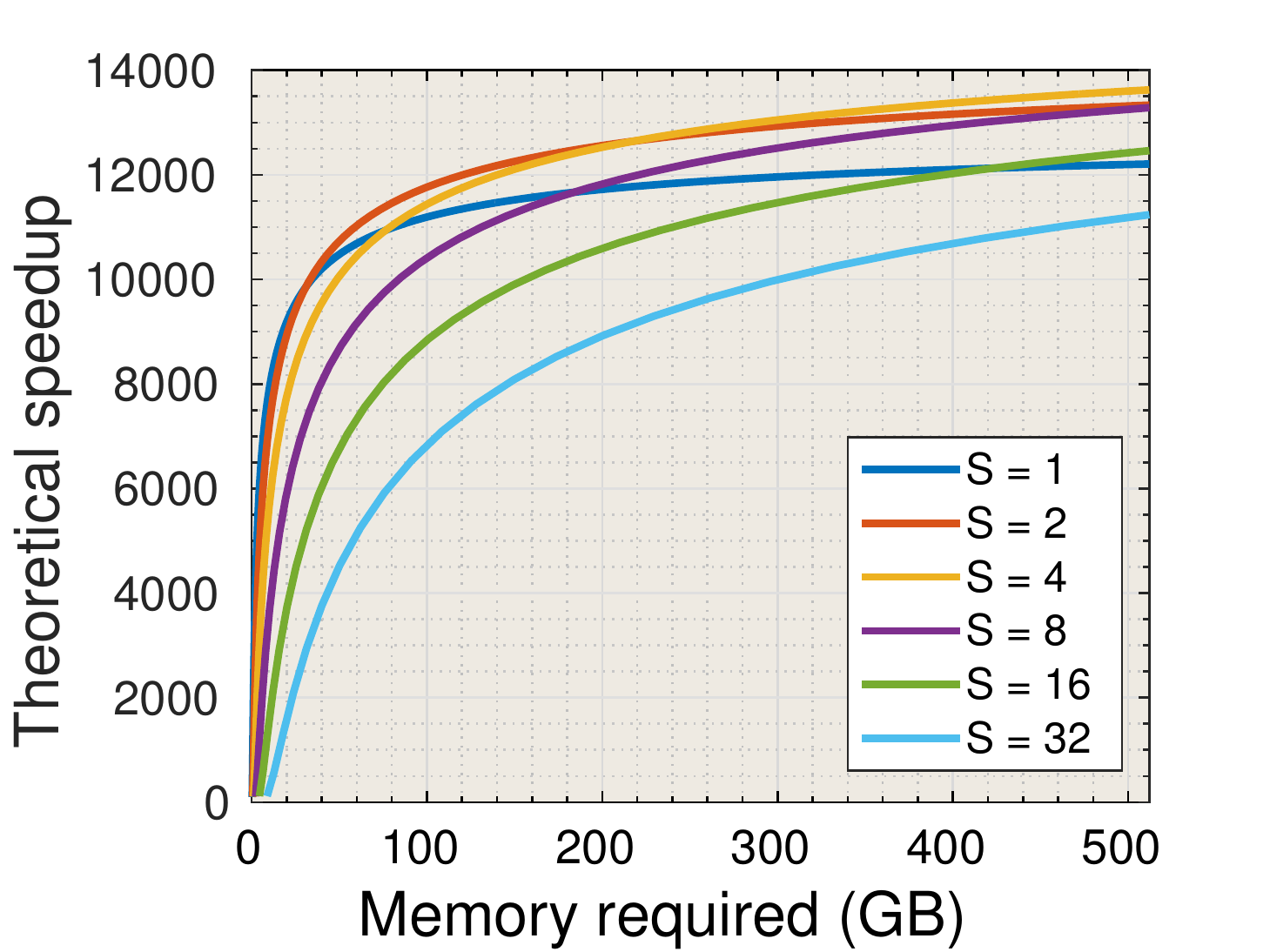}
      \label{fig:fftbatcha}
    }
    \subfloat[]{\protect\includegraphics[trim= 0mm 0mm 9mm 1mm, clip, width=0.5\columnwidth]
      {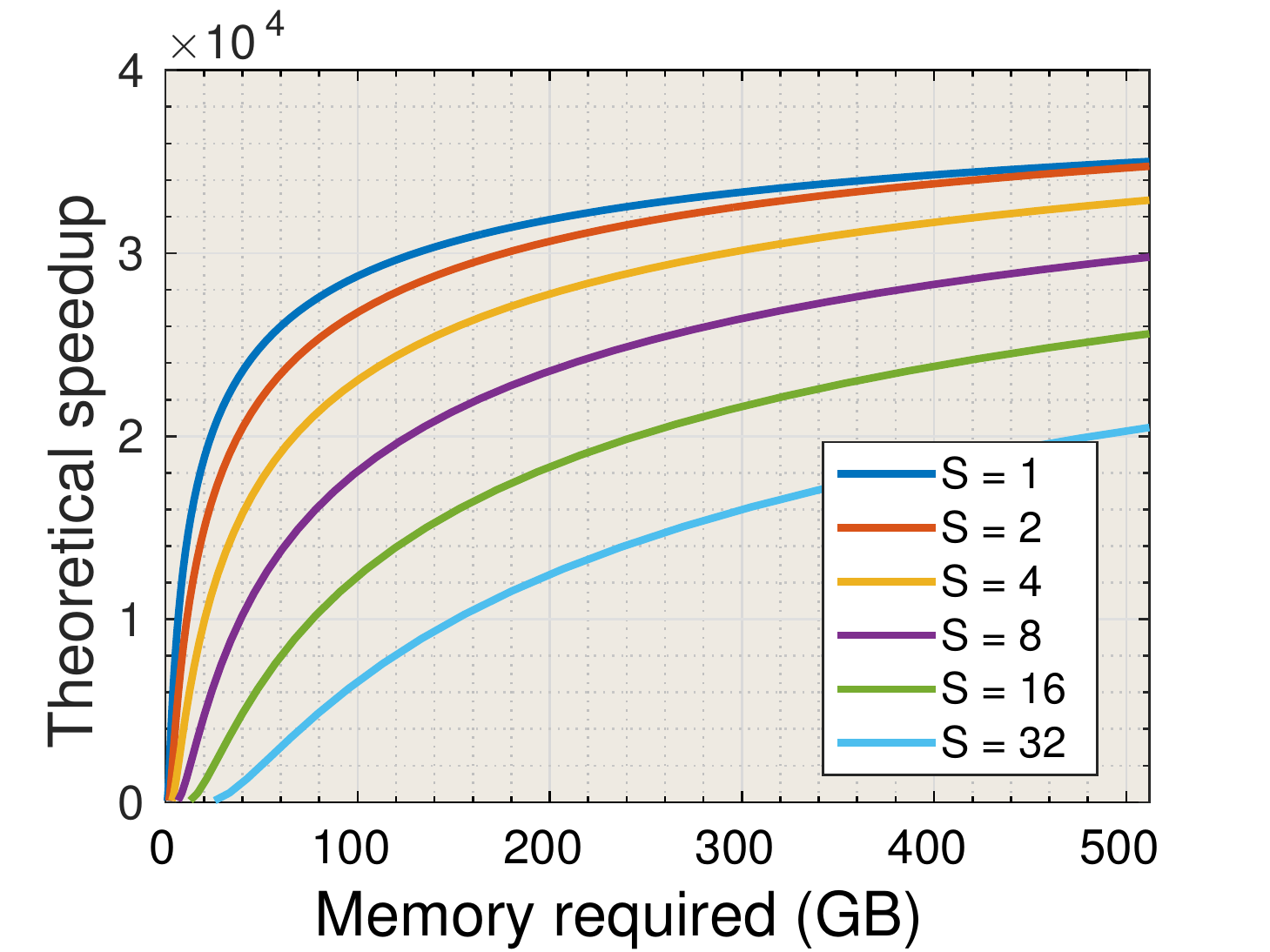}
      \label{fig:fftbatchb}
    }
    \caption{Theoretical speedup of pooling networks using FFT--based
      convolution for different sizes of the inputs and different
      batch sizes for a network with 1 pooling layer (a) and 2 pooling
      layers (b).}
    \label{fig:fftbatch}
  \end{figure}

  %

  \begin{figure*}[!htbp]
    \centering 
    \subfloat[]{\protect\includegraphics[trim= 6mm 0mm 9mm 1mm, clip, width=0.25\textwidth]
      {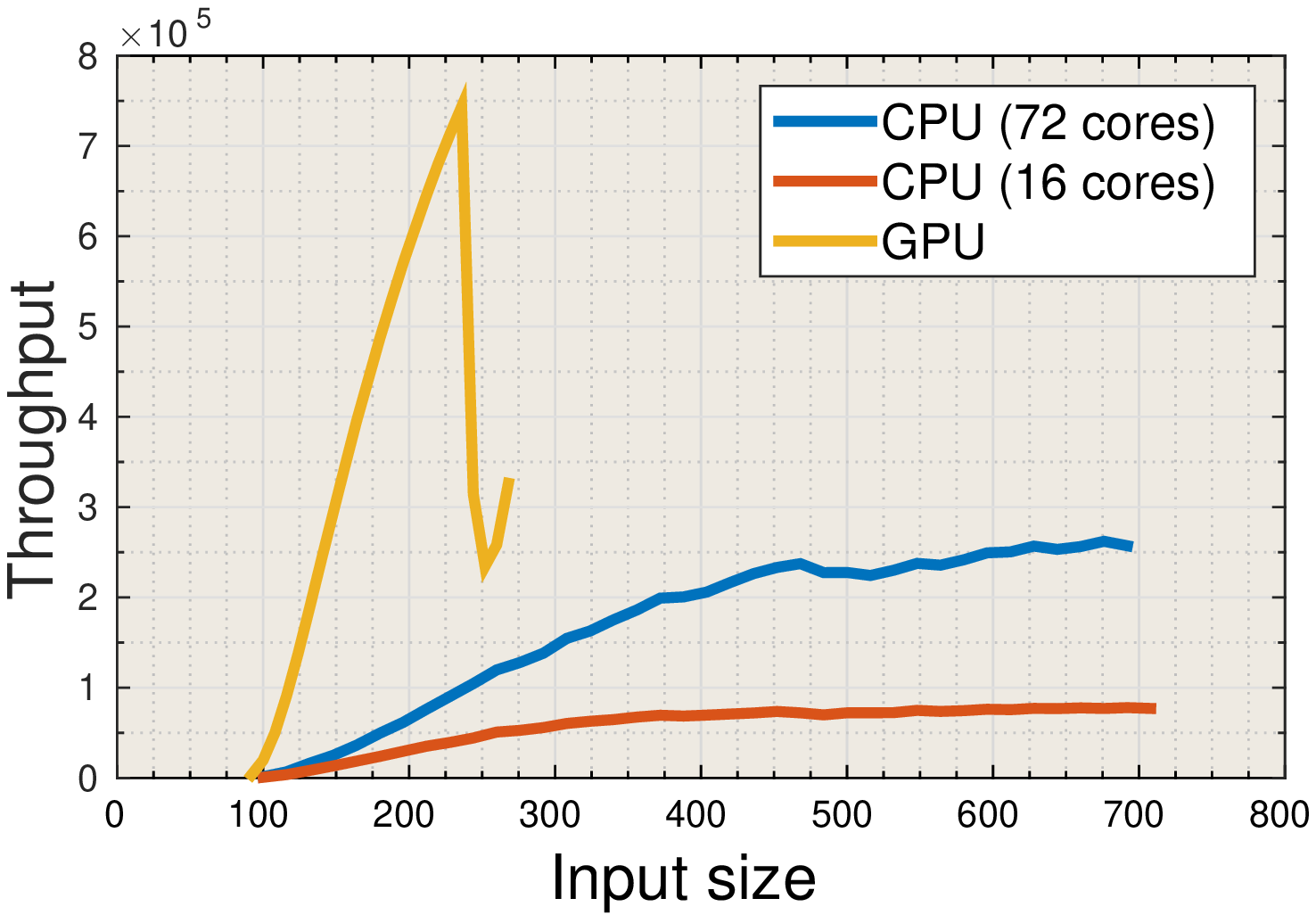}}
    \subfloat[]{\protect\includegraphics[trim= 6mm 0mm 9mm 1mm, clip, width=0.25\textwidth]
      {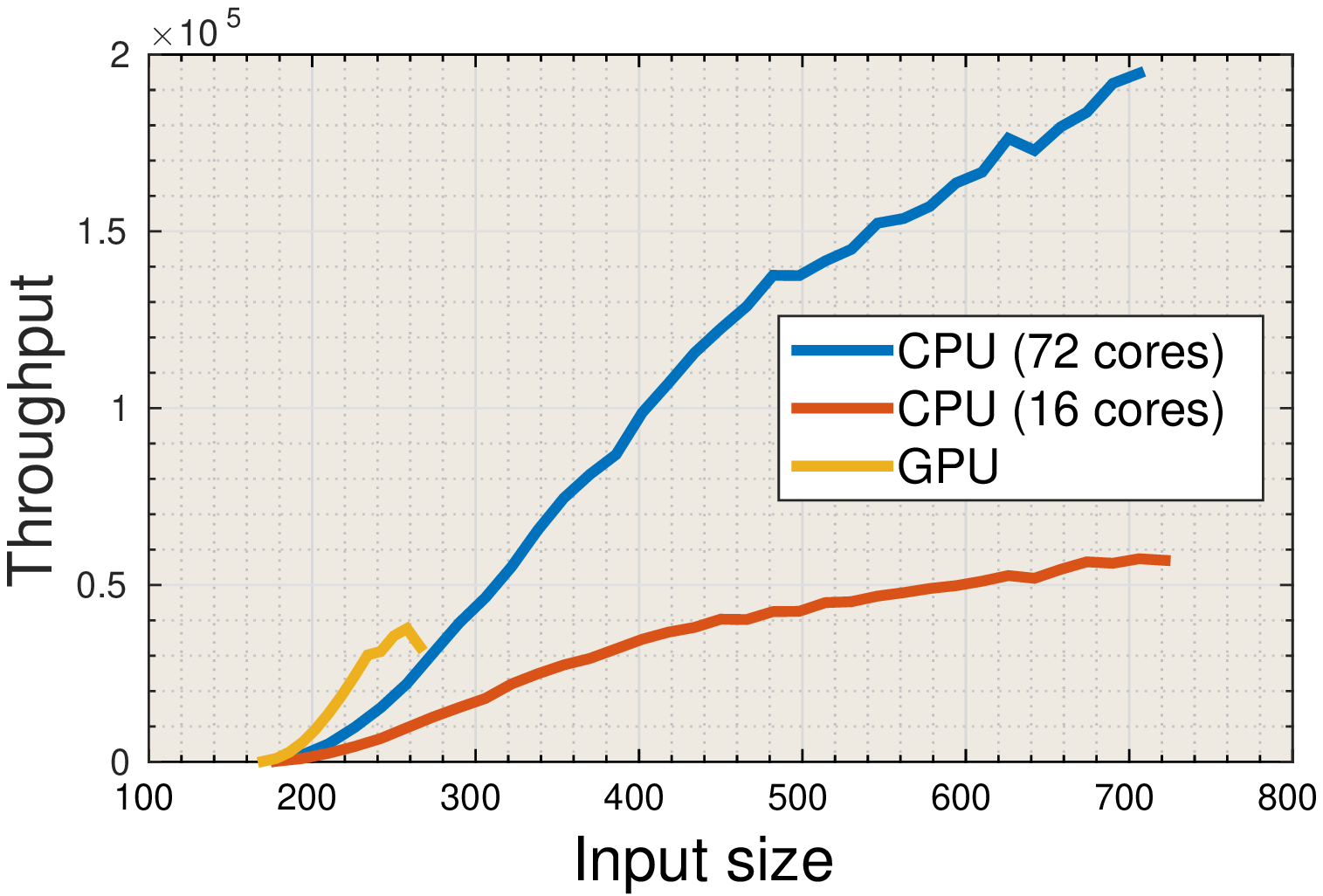}}
    \subfloat[]{\protect\includegraphics[trim= 6mm 0mm 9mm 1mm, clip, width=0.25\textwidth]
      {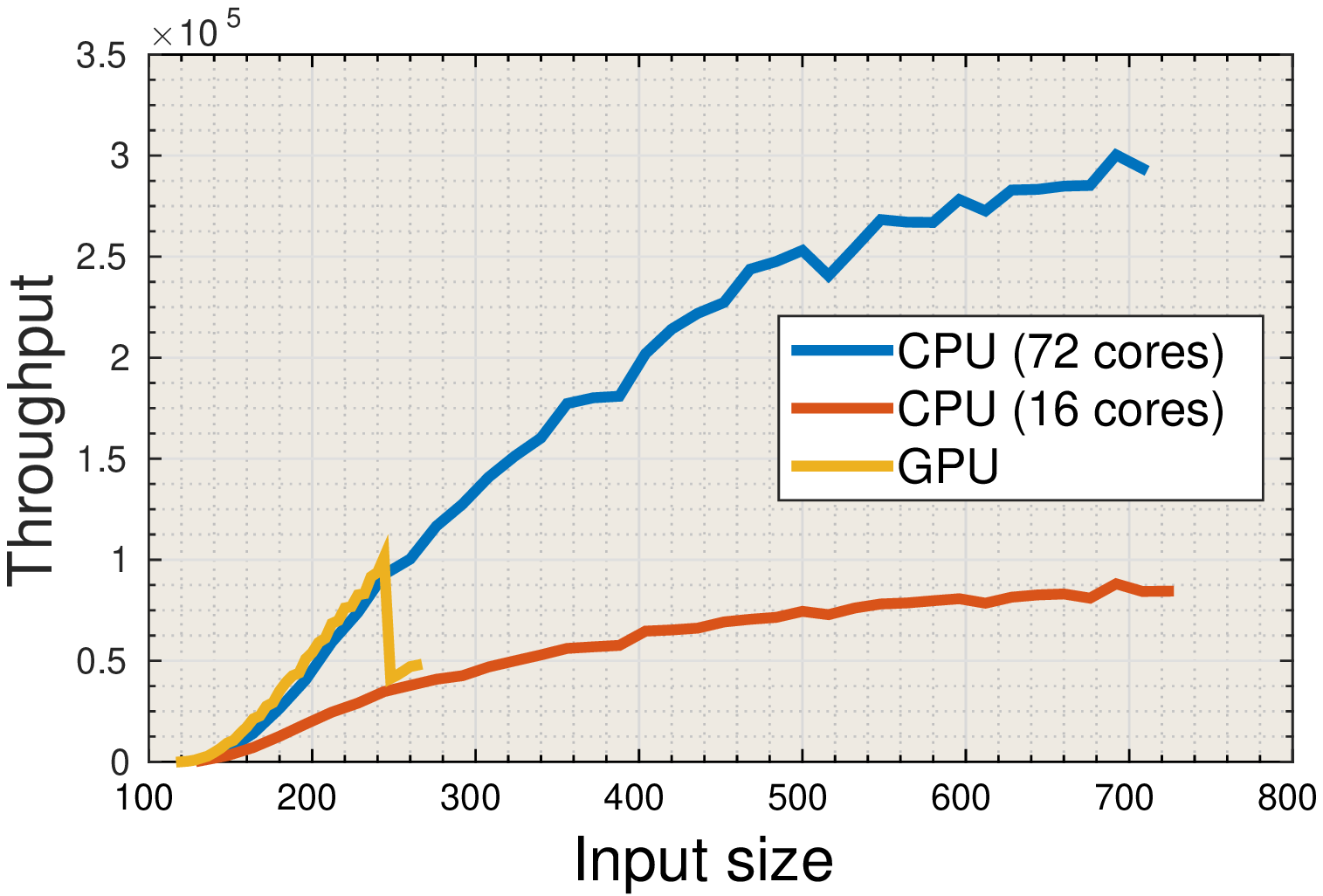}}
    \subfloat[]{\protect\includegraphics[trim= 6mm 0mm 9mm 1mm, clip, width=0.25\textwidth]
      {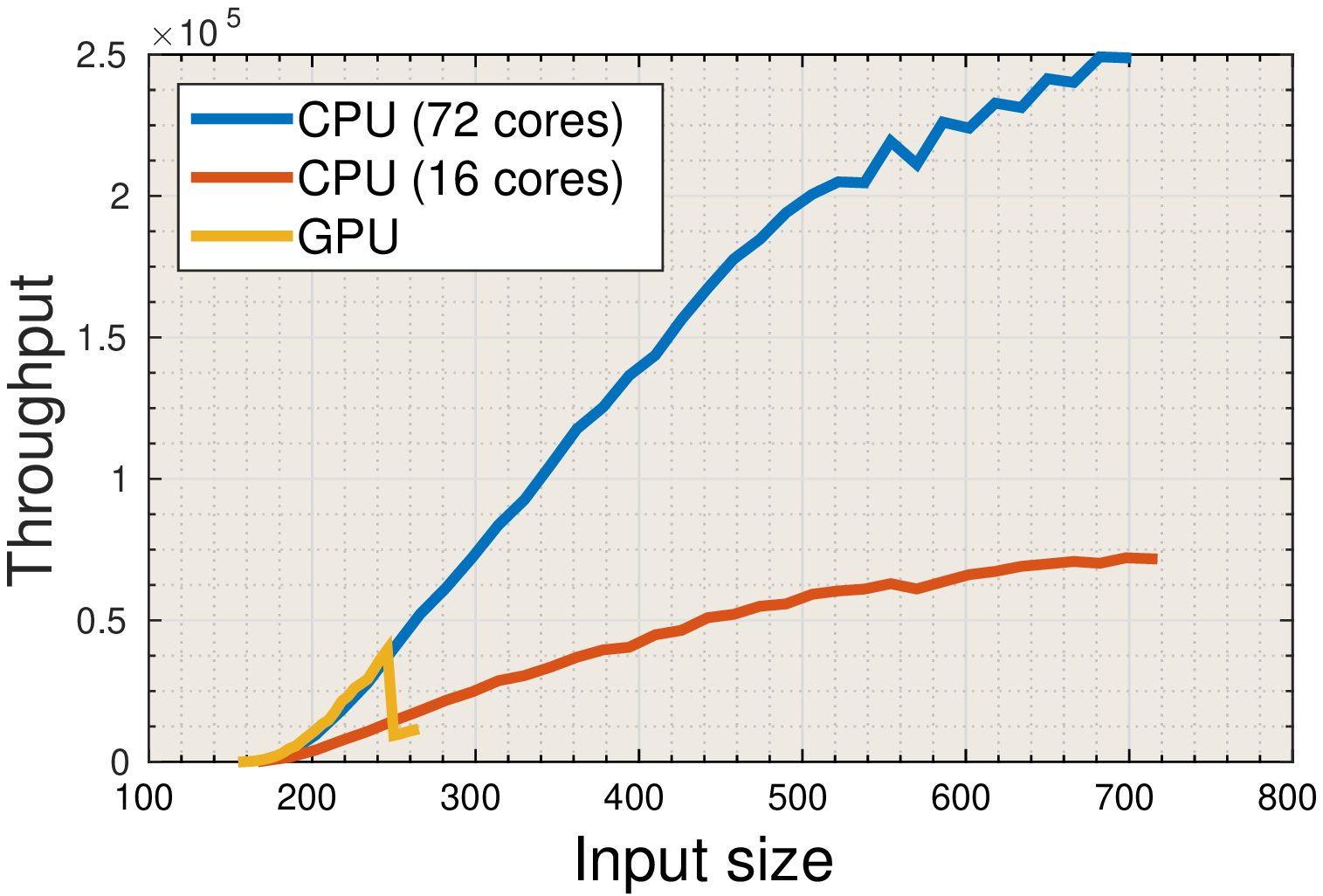}}

    \caption{Maximal throughput achieved vs input image size using
      GPU--only and CPU--only primitives.}
    \label{fig:experiments1}
  \end{figure*}

  The fact that $S=1$ produces better results is more surprising, and
  we attempt to give some theoretical intuition here.  Increasing the
  batch size can greatly reduce the computational cost of FFT--based
  ConvNets by reusing the transforms of the
  kernels~\cite{mathieu-iclr-14,vasilache2014fast}.  However, when the
  batch size is larger, the memory requirement for each layer
  increases, thus effectively limiting the maximal size of the input.
  Increasing the batch size, as well as the input image size increase
  the memory requirements of the algorithm and lower the computational
  complexity when FFT based convolutions are used.  In our case,
  increasing the image size had a stronger effect, which is not always
  the case.

  On Fig.~\ref{fig:fftbatch} we show the theoretical speedup
  achievable by using different batch sizes for two networks, using
  the computational cost of an FFT--based layer shown in
  Table~\ref{table:layers_complexity}.  The theoretical speedup is
  defined as the ratio of operations required to compute a single
  output pixel of a naive approach (having input image equal to the
  field of view of the network -- output equal to $1 \times 1 \times
  1$, and a network using \emph{MPF} with different input sizes.
  Instead of having the input size on the $x$ axis we display the
  memory required for such network.

  We see that for a network with 2 pooling layers, having a batch size
  of $1$ can achieve the highest theoretical speedup.  As the networks
  we benchmark have 2 or more pooling layers our empirical results
  agree with the theoretical estimate.  When optimizing the throughput
  of networks with only 1 pooling layer, one should consider larger
  batch sizes.

\subsection{Empirical results}
  We expect that algorithm throughput will depend on ConvNet
  architecture.  For benchmarking purposes, we chose to make the field
  of view fairly large.  This was accomplished either by having more
  pooling layers or larger filters. Two architectures had seven
  convolutional and three pooling layers in the sequence CPCPCPCCCC.
  The networks had kernel sizes of $3\times 3 \times 3$ and $5\times
  5\times 5$.  Two other architectures had 6 convolutional and 2
  pooling layers (CPCPCCCC) with larger filter sizes of $7 \times
  7 \times 7$ and $9 \times 9 \times 9$.  The architectures of all
  four benchmarked networks are shown in
  Table~\ref{table:benchmarked_networks}.  A rectified linear transfer
  function is applied after each convolutional layer.  The complexity
  of the transfer function has little influence on computation
  time as it represents only a small fraction of the overall
  computation.

  The benchmarks are performed on two machines.  The first machine is
  a 4-way Intel Xeon E7 8890v3 with total of $72$ cores ($144$
  hyper--threads), 256GB of RAM and a Titan X GPU (with 12GB on--board
  RAM).  The second machine is an Amazon EC2 instance with $32$
  virtual cores and 244GB of RAM (r3.8xlarge).  The second machine is
  included as it is more readily available.

  Fig.~\ref{fig:experiments1} shows the throughput achieved on the
  four benchmarked networks (Table~\ref{table:benchmarked_networks})
  with the batch size $S=1$, and different image sizes.

  It turns out that the optimal choice for primitives for the CPU is
  always the same regardless of the network choice.  In all cases the
  first (convolutional) layer was optimized to data--parallel
  FFT--based algorithm, and the rest of the convolutional layers used
  the task--based algorithm.  This is expected because of the much
  higher cache locality of our FFT implementations compared to direct
  convolution, which plays an important role on the CPU with very fast
  cache access and relatively slow RAM access.


  \begin{table}
    {\footnotesize
    \centering
    \begin{tabular}{ccccc}
      \toprule
      & n337    & n537  &  n726  &  n926 \\
      \midrule
      Input size & $235^3$ &  $217^3$  & $243^3$  & $253^3$ \\
      \midrule
      Layer $1$  &  CuDNN1  &  CuDNN1  & CuDNN1   & CuDNN1  \\
      Layer $2$  &  MPF     &  MPF     & MPF      & MPF     \\
      Layer $3$  &  CuDNN2  &  FFT     & FFT      & CuDNN1  \\
      Layer $4$  &  MPF     &  MPF     & MPF      & MPF     \\
      Layer $5$  &  CuDNN2  &  CuDNN2  & FFT      & FFT     \\
      Layer $6$  &  MPF     &  MPF     & FFT      & FFT     \\
      Layer $7$  &  CuDNN2  &  CuDNN2  & FFT      & FFT     \\
      Layer $8$  &  CuDNN2  &  CuDNN2  & FFT      & FFT     \\
      Layer $9$  &  CuDNN2  &  CuDNN2  & & \\
      Layer $10$ &  CuDNN2  &  CuDNN2  & & \\
      \bottomrule
    \end{tabular}
    \caption{Optimal choice for different layers.}
    \label{table:gpu_optimal}
    }
  \end{table}

  The optimal use of primitives for the GPU have higher dependence on
  the ConvNet's architecture.  The optimal choice of primitives for
  each network, as well as the optimal input image size is given in
  Table~\ref{table:gpu_optimal}.  Interestingly, the implementation
  for the first layer of all networks is the slower version of the
  cuDNN's primitive.  Even though the primitive is slower, it is able
  to process larger input image, as it has lower memory requirement.
  There is a trade--off between the layer speed, and the maximal size
  of the input that a layer can process.  In this case it was more
  beneficial to be able to process larger inputs.

\section{GPU + host RAM and CPU--GPU inference}

  We saw that the throughput can highly depend on the size of the
  input image we are able to process by the network.  For kernels of
  size $5^3$ or more CPU seems to outperform GPU due to these
  limits.  For that reason we introduce another layer primitive, a
  layer whose data is stored in host RAM, and is partially uploaded to
  the GPU where the computation is performed.

\subsection{GPU + host RAM convolutional layer}

  \begin{figure}
    \centering
    \subfloat[]{\protect\includegraphics[width=0.07\textwidth]
      {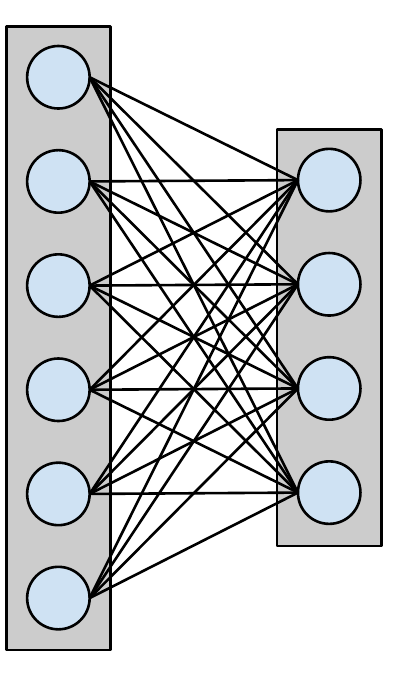}
      \label{fig:partial_execa}
    }
    \subfloat[]{\protect\includegraphics[width=0.3\textwidth]
      {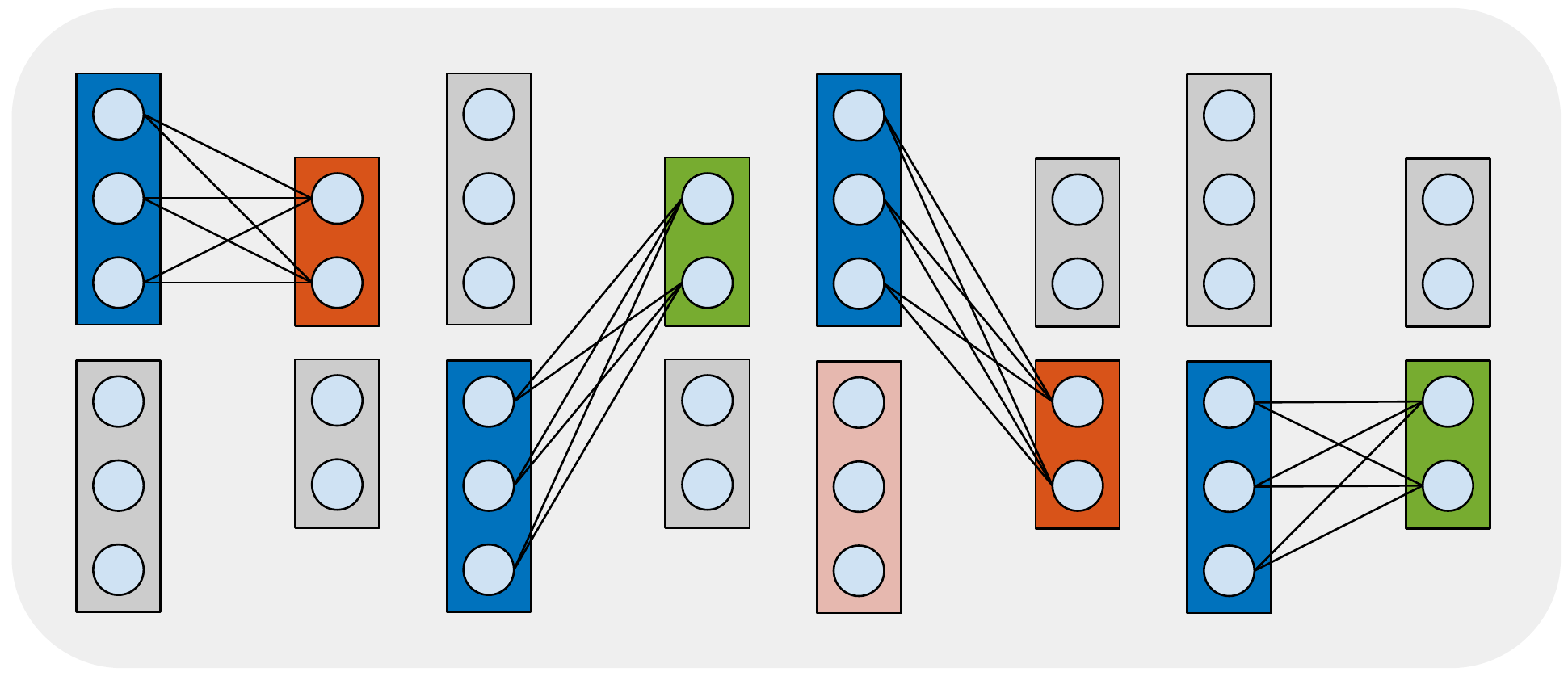}
      \label{fig:partial_execb}
    }
    \caption{Decomposing a convolutional layer into multiple
      convolutional sub--layers.}
    \label{fig:partial_exec}
  \end{figure}

  Consider a convolutional layer whose input is shape is $(S,f,x,y,z)$
  and output shape $(S,f',x',y',z')$.  The computation performed by
  the layer can be divided into $N$ sub--layers with input shapes of
  $(S_i,f_i,x,y,z)$ and output shape $(S_i,f'_i,x',y',z')$.
  Fig.~\ref{fig:partial_exec} illustrates how the computation of a
  convolutional layer with $S=1$, $f=6$, and $f'=4$ can be divided
  into $N=4$ sub--layers, each having $S_i=1$, $f_i=3$ and $f'_i=2$.
  The blue color represents the input images that have to be
  transferred to the GPU, the red color represents the memory that has
  to be allocated on the GPU.  The green color represent the results
  that have to be transferred back to the host.  The computation of
  each sub--layer can be performed by any of the GPU--only primitives.
  The time required for processing the layer will equal to the sum of
  processing time of each sub--layer and the time required for memory
  transfers.  Due to the GPU's memory limit, not all divisions are
  feasible.

  Due to a large number of possible divisions, simple way to find the
  optimal division into sub--layers by considering all feasible
  divisions, and then picking the optimal primitive for each
  sub--layer might be very time consuming.  The search can be pruned
  using the following two heuristics.

  First, for small kernel sizes ($5^3$ or smaller), we consider only
  the cuDNN primitives based on direct--convolution; and for larger
  kernels we only consider our FFT--based primitive.

  Second, when $S > 1$, we prefer the division into sub--layers such
  that $f_i=f$, $f'_i=f'$, and $S_i \le S$.  This, essentially divides
  the batches of the input into sub--batches that are processed on the
  GPU.  In this case, each input will have to be transferred to the
  GPU exactly once, and each output will be transferred back to the
  host exactly once.

  If there is no feasible division into sub--batches, we prefer
  sub--layers with $S_i = 1$, where most of the sub--layers have $f_i
  = f_{\alpha} \le f$ and $f'_i = f'_{\alpha} \le f'$.  Some
  sub--layers will have to have a different input/output shape if $f$
  or $f'$ are not divisible by $f_{\alpha}$ and $f'_{\alpha}$,
  respectively.  As most of the sub--layers have the same input and
  output shape, the time required for the layer can be estimated by
  running only sub--layers with distinctive input shapes.

  The two heuristics render the search for the optimal sub--division,
  and optimal choice for each sub--layer very fast.

  The host memory requirement for this layer equals to the amount of
  memory required to store the input and the output tensor, and GPU
  on--board memory has to be large enough to facilitate each
  sub--layer.

\subsection{GPU + host RAM ConvNet execution}

  Implementing GPU + host RAM MPF layer turned out to be impractical.
  Originally we implemented similar GPU + host RAM MPF layers.
  However, it turned out that it is better to compute the MPF layers
  on the CPU, even when very few cores are available.  This is because
  of the expensive transfer to and from device and relative low
  computational complexity of the MPF layers.

  \begin{figure*}[!htbp]
    \centering
    \subfloat[]{\protect\includegraphics[trim= 6mm 0mm 9mm 1mm, clip, width=0.25\textwidth]
      {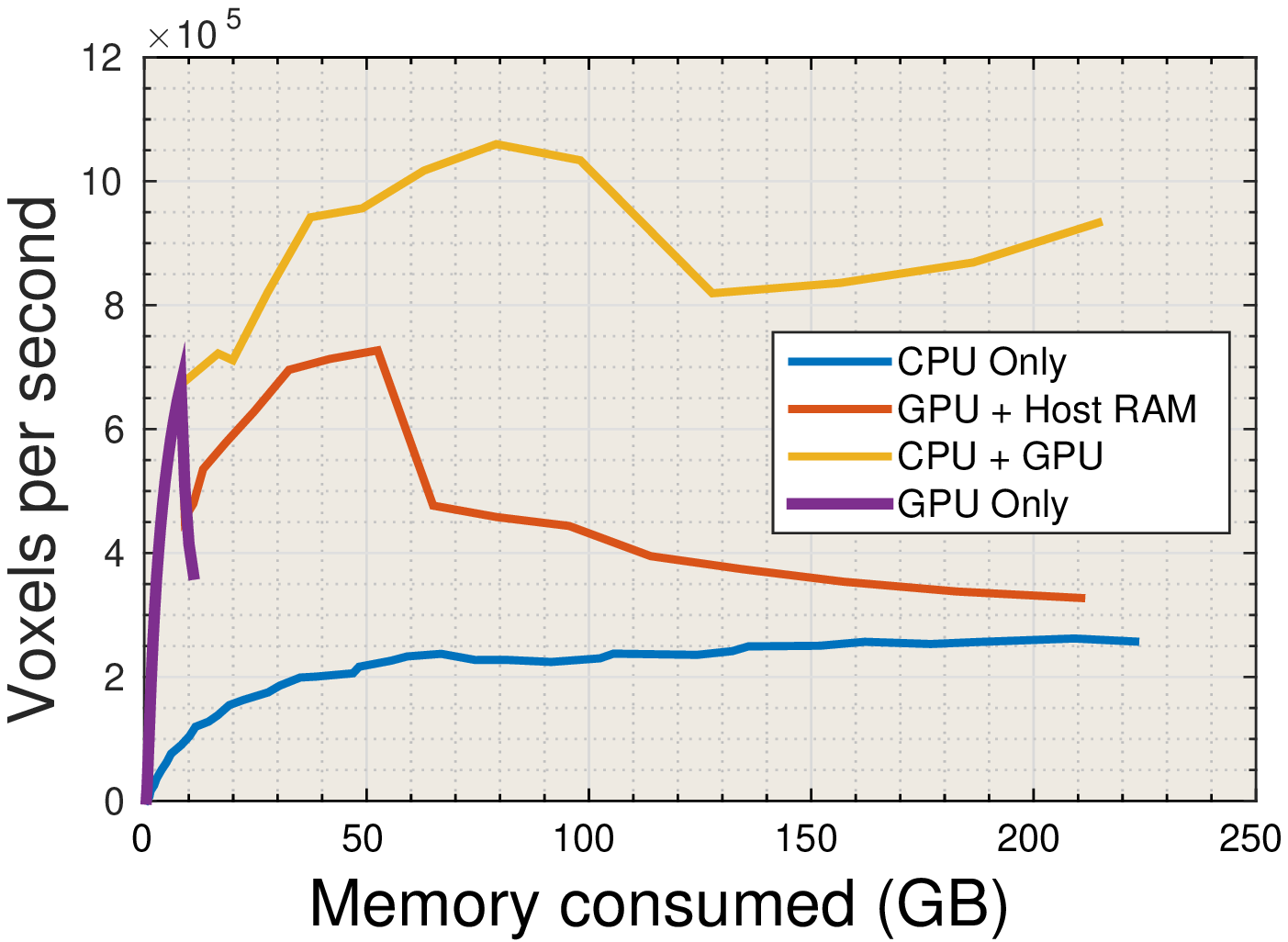}}
    \subfloat[]{\protect\includegraphics[trim= 6mm 0mm 9mm 1mm, clip, width=0.25\textwidth]
      {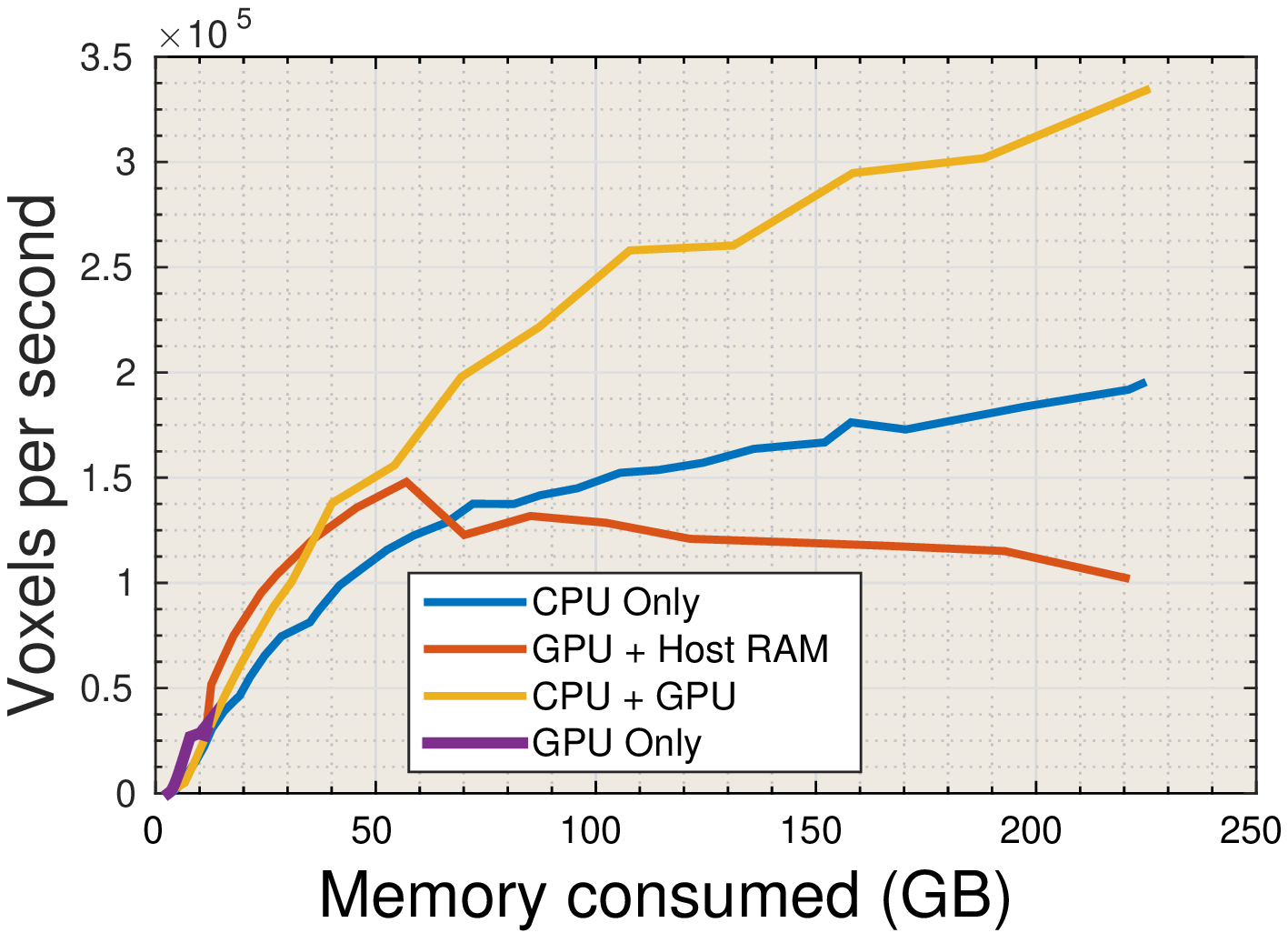}}
    \subfloat[]{\protect\includegraphics[trim= 6mm 0mm 9mm 1mm, clip, width=0.25\textwidth]
      {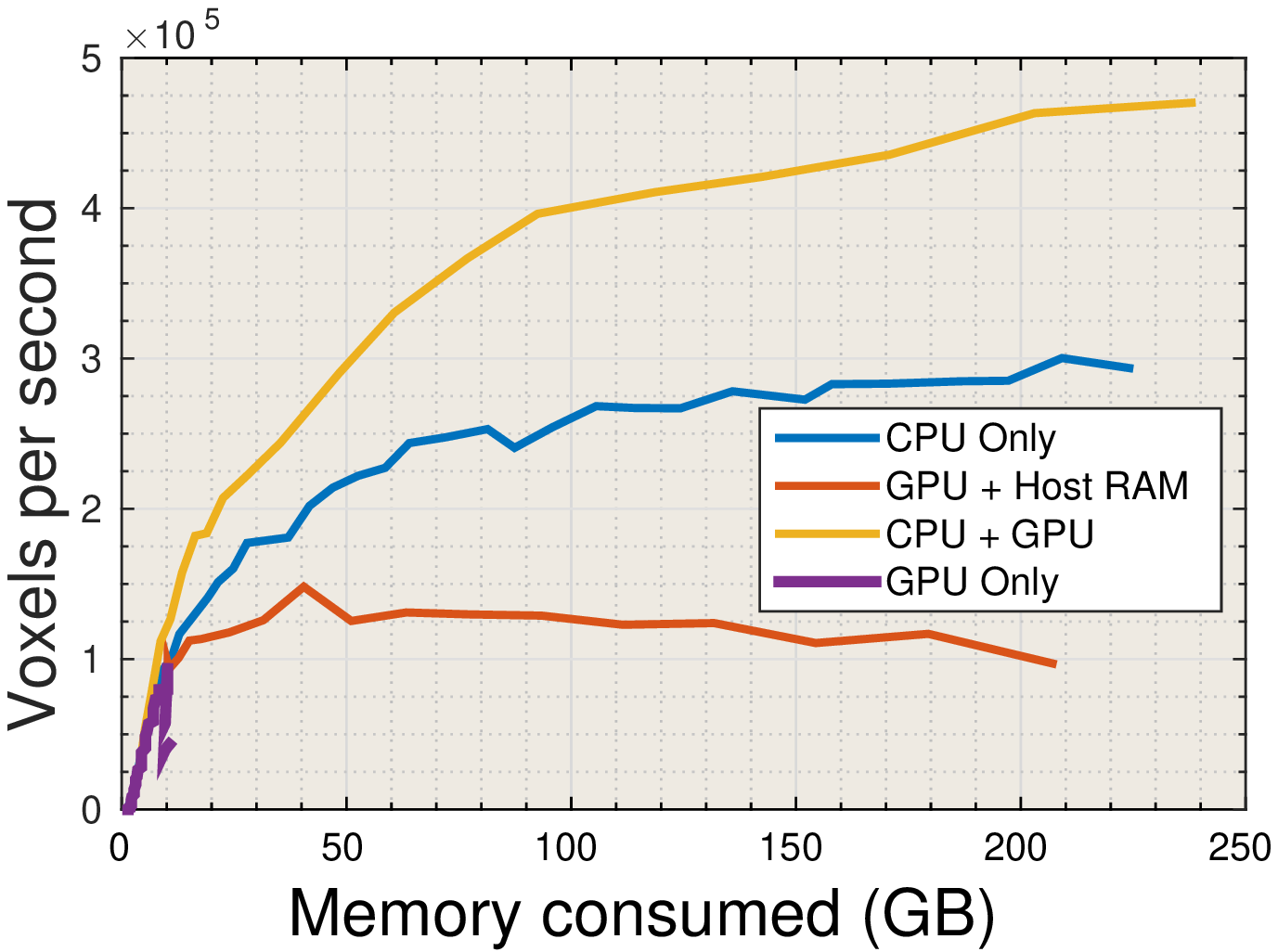}}
    \subfloat[]{\protect\includegraphics[trim= 6mm 0mm 9mm 1mm, clip, width=0.25\textwidth]
      {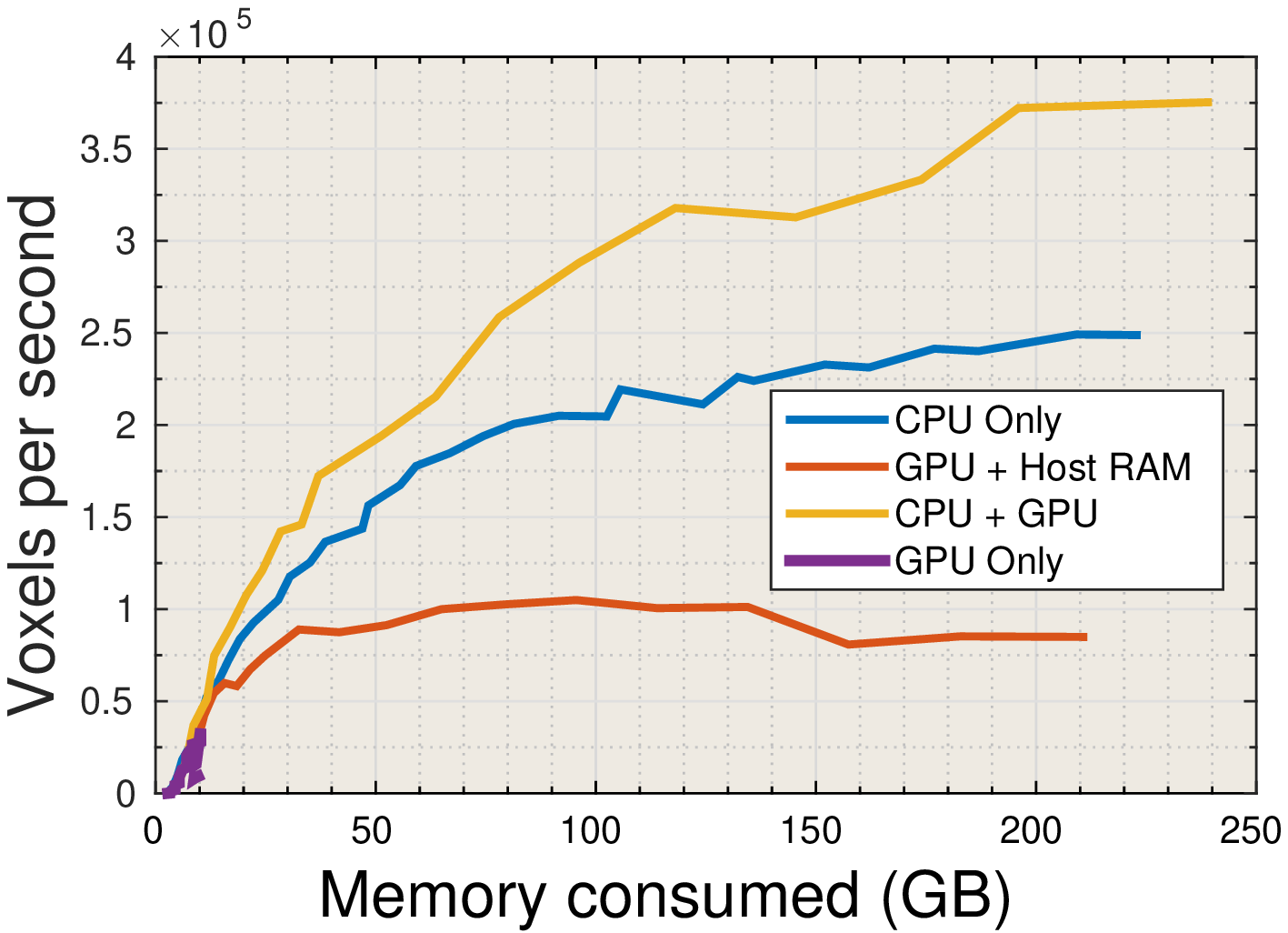}}

    \caption{Maximal throughput achieved vs memory consumption using
      GPU--only, CPU--only, CPU + host RAM and CPU--GPU
      implementations for different image sizes.}
    \label{fig:final_results}
  \end{figure*}

  \begin{figure}
    \begin{center}
      \includegraphics[width=0.99\columnwidth]{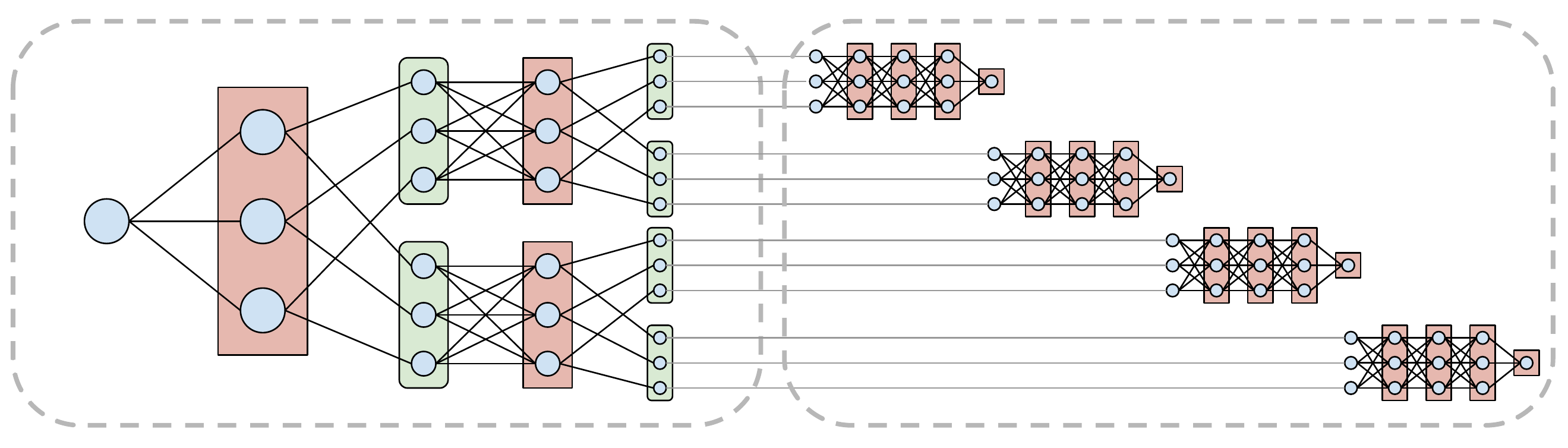}
    \end{center}
    \caption{A network of form CPCPCCCC with the first half executed
      one layer at the time, and the second half one batch at the
      time.  We pooling window size of the MPF layers is $2 \times 1
      \times 1$. }
    \label{fig:layer-vs-batch}
  \end{figure}

  The simplest way to execute a network using the GPU for computation
  and host RAM for storage is to use individually optimized GPU + host
  RAM layer described above for each convolutional layer, and the CPU
  implementation of a MPF layer for each pooling layer.

  Turns out that, when using MPF layers for evaluating a max--pooling
  ConvNet we can achieve better performances by minimizing the data
  transferred to and from the GPU.  To understand how we can improve
  the performances, we need to understand the following property of a
  ConvNet.

  A ConvNet with input shape $I = (S,f,x,y,z)$ with $S > 1$, will have
  the batch size $S'$ of the output shape $I' = (S',f',x',y',z')$
  always divisible by $S$.  The values $I'(\frac{S'}{S} i :
  \frac{S'}{S} (i+1),:,:,:,:)$ will only depend on $I(i,:,:,:,:)$.
  This means that concatenating results of applying a layer on two
  inputs $I_1(S_1,f,x,y,z)$ and $I_2(S_2,f,x,y,z)$ will equal to the
  result of the concatenated input of size $(S_1+S_2,f,x,y,z)$.

  \begin{table*}[!htbp]
    { \footnotesize
    \centering
    \begin{tabular}{l|cccccccc}
      \toprule
      Network  & Baseline (cuDNN)  & Caffe    & ELEKTRONN   & ZNN         & GPU-Only    & CPU-Only    & GPU + host RAM & GPU-CPU   \\
      \midrule
      n337     & $22,934.8$        & $1.348$  & $122,668$   & $34,334.8$  & $671,782$   & $262,131$   & $727,103$      & $1,059,910$ \\
      n537     & $1,048.68$        & --       & --          & $9,494.5$   & $29,352.1$  & $194,683$   & $147,965$      & $334,163$ \\
      n726     & $13,520.4$        & --       & $6,122$     & $31,354.8$  & $97,257.2$  & $300,312$   & $148,194$      & $470,166$ \\
      n926     & $2,667.86$        & --       & --          & $20,908.6$  & $35,051.3$  & $249,190$   & $104,946$      & $375,295$ \\
      \bottomrule
    \end{tabular}
    \caption{Comparisons to other methods.}
    \label{table:comparisons_to_others}
  }
  \end{table*}

  Consider the output shape of the first $\theta$ layers of a ConvNet.
  The computation of the rest of the layers can be considered to be
  another ConvNet that takes the output of the $\theta^{th}$ layer as
  the input.  If some of the first $\theta$ layers were MPF layers,
  the batch size of the $\theta^{th}$ layer output $S_{\theta}$ will
  be greater than 1.  Instead processing one layer at the time for the
  rest of the layers, one might be able to process all remaining
  layers for a sub--batch $\hat{S}_{\theta}$ at the time using a
  GPU--only network.  This will reduce the memory transfer overhead as
  no intermediate results have to be transferred back to the host.  In
  Fig.~\ref{fig:layer-vs-batch} we illustrate the timeline of
  executing a network of the form CPCPCCCC by having the first four
  layers executed one layer at the time, and the rest one batch at the
  time.

We illustrate the stragety through
  the follwing example.  Assume we are given a ConvNet architecture in
  the form CPCPCCCC, where each layer has 3 input and 3 output images,
  except for the first layer that has 1 input image and the last that
  has 1 output image.  Let the sizes of the pooling window be $2
  \times 1 \times 1$, this means that when \emph{MPF} layers is used,
  for each input tuple a \emph{MPF} layer will produce 2 fragments of
  the same size.  Mainly the most significant dimension of the input
  tensor to the layer will double.  If the batch size $S$ of the
  network input was $1$, the output of the second \emph{MPF} layer
  will have the batch size of $4$.  At this point we might be able to
  process the rest of the network for a single input tuple (or
  multiple) using the GPU--only approach, thus only having to upload
  the input to the GPU once, and fetch the final result once.
  Processing the first four layers of the network using the GPU + host
  RAM and MPF layers and the rest of the layers with the GPU--only
  approach for the rest of the ConvNet is illustrated in
  Fig.~\ref{fig:layer-vs-batch}.  Here we process the first four
  layers, one layer at the time, and the last four layers one batch at
  the time.

  Finding the optimal network execution strategy for such execution
  becomes more complex.  The first additional parameter we have is
  $\theta$, $(0 \le \theta \le L)$, where $L$ is the number of layers
  of the given ConvNet.  This parameters represent the number of
  layers that will be processed one at the time using the GPU + host
  RAM or CPU-MPF layer at the time.  The rest of the network is then
  executed one (or more) batches at the time using the GPU--only
  primitives.

  For a given value of $\theta$ and a given input size, this approach
  has two limitations.  Firstly, $\theta$ layers have to fit on the
  host RAM.  Secondly, there has to exist a GPU--only network that can
  process the latter layers on the GPU.

  In order to find an optimal implementation we, consider any valid
  input shape and any valid $\theta$ that are convolutional to use GPU
  + host RAM primitive, and separately optimize the rest of the
  GPU--only network

\subsection{CPU--GPU ConvNet execution}

  Finally, inference can be done by utilizing both CPU and GPU.  As in
  the GPU + host RAM approach, the network layers are divided into two
  groups.  For the first $\theta$ layers, we use the optimal CPU
  implementation as defined in Section VI, and for the rest of the
  layers we use the optimal GPU implementation as defined in the
  previous section.

  The CPU and the GPU form a producer--consumer pipeline.  The CPU
  produces by computing the first $\theta$ layers for a given input
  image, and queuing the result.  The GPU consumes the data on the
  queue, taking as input the output of the $\theta^{th}$ layer, and
  yields the final output of the last layer.

  This approach can generate huge memory overhead if the CPU produces
  data much faster than the GPU can consume.  For that reason, the CPU
  is not allowed to start working on the next input until the queue is
  empty -- until the GPU had picked up and started executing the rest
  of the network for all the data on the queue.  This essentially
  limits the queue to a maximal size of one.

  For a given value of $\theta$ and a given input size the GPU will
  operate on the output of the $\theta^{th}$ layer producing the final
  output.  Hence the output of the $\theta^{th}$ layer has to be
  stored in the host RAM, along with memory allocated for the network
  output.  As both the CPUs and GPUs can do work only when the queue
  is empty, the rest of the host RAM is available to the CPUs.

  Finding the optimal implementation through an exhaustive search
  resembles the one in the previous section.  For each valid input
  shape, we loop over all valid values of $\theta$, and for each such
  division of the ConvNet, we separately optimize the first $\theta$
  CPU-only layers and the rest of the GPU-only layers, having the
  memory limitations in mind.

  Fig.~\ref{fig:final_results} shows the throughput achieved on the
  four benchmarked networks (Table~\ref{table:benchmarked_networks})
  with the batch size $S=1$, and different image sizes obtained during
  the exhaustive search for the highest throughput.  Instead of
  showing the input image size on the $x$ axis, we decide to show the
  memory required by the implementation, as it allows us to analyze
  how the memory available to the system influences the maximal
  possible throughput.  The memory consumed is calculated as
  $\max\{M_{CPU},M_{GPU}\}$.

\section{Comparison to other algorithms}

  We compare our 4 approaches (GPU--only, CPU--only, GPU + host RAM
  and GPU--CPU) with other publicly available implementations and show
  the results in the Table~\ref{table:comparisons_to_others}.  All
  benchmarks are performed on the same hardware, the 4--way Intel Xeon
  E7–-8890 v3 machine with 256GB of RAM and a Titan X GPU.

  The results of our approaches are based on the optimizations
  described in the previous sections.  For the other approaches, we
  varied the input sizes and measured the throughput, we reported the
  highest value of throughput obtained, which was always correlated
  with the size of the input we were able to process.

  The baseline (cuDNN) approach consists of calling the
  cuDNN~\cite{chetlur2014cudnn} primitives for convolution and
  max--pooling.  Unlike other approaches, this is not a general
  framework -- it requires the user to write some code for calling
  into the cuDNN primitives, for calculating the input/output shapes
  and some minimal memory management.  We expect that a user with
  minimal programming experience could implement the above.  Our
  implementation was done in C++, however one could use cuDNN bindings
  for other languages.

  Caffe~\cite{jia2014caffe} is another GPU ConvNet framework.  We
  benchmarked a version that implements sliding window ConvNets using
  ``strided kernels''~\cite{tschopp2015efficient}.  The implementation
  is also optimized for training and seems to have a huge memory
  overhead as we were only able to run the smallest of the networks.

  ELEKTRONN~\cite{ELEKTRONN2015} was the only competitor that provides
  inference optimization for 3D sliding window ConvNets using
  \emph{MPF}.  The package also uses cuDNN convolutional primitives.
  However, it was able only to process two of our four networks.

  ZNN~\cite{zlateski2015znn} is a framework optimized for training
  sliding window 3D ConvNets on multi--core and many--core CPUs using
  ``max--filtering'' followed by FFT--based ``sparse convolution''.
  ZNN was the best competitor for networks with filters of $5 \times 5
  \times 5$ or larger.

  Our CPU-only, GPU-only, GPU + host RAM, and CPU-GPU implementations
  outperform all competitors.  For the smallest ConvNet architecture,
  {\tt 237} with the kernel sizes of $3^3$, the next best competitor
  was ELEKTRONN with approximately a tenth of the speed of our
  CPU--GPU approach. For all the other ConvNets, the next best
  competitor was ZNN with approximately $15 \times$ smaller
  throughput.



\section*{Acknowledgments}
We thank Kai Li and Nir Shavit for helpful discussions.  We are
grateful to Intel Corporation for providing the 4–-way Intel Xeon
E7–-8890 v3 machine, and for supporting the Intel Parallel Computing
Center at Princeton University.  We acknowledge support from IARPA
(D16PC00005), the Mathers Foundation, NIH/NINDS, and the U.S. Army
Research Office (W911NF-12-1-0594). Kisuk Lee was supported by a
Samsung Scholarship.



\end{document}